 \documentclass[amssymb,12pt,showpacs,footinbib]{revtex4}
\usepackage[]{graphicx,amsmath}
\usepackage{color}
\usepackage[all]{xy}
\usepackage{amsthm}
\usepackage{pdfsync}

\usepackage[pdftex]{hyperref}

\def\ket#1{| #1 \rangle}

\newcommand {\be} {\begin{eqnarray}}
\newcommand {\ee} {\end{eqnarray}}
\newcommand {\bp} {\begin{pmatrix}}
\newcommand {\ep} {\end{pmatrix}}
\newcommand{\ham}{\mathcal{{H}}}

\newcommand{\U}{\hat{U}}
\newcommand{\R}{{R}}

\newcommand{\sx}{{\sigma}_x}
\newcommand{\sy}{{\sigma}_y}
\newcommand{\sz}{{\sigma}_z}

\newcommand{\B}{\vec{B}_1}
\newcommand{\rpsi}{\ket{\psi}_r}
\newcommand{\up}{\uparrow}
\newcommand{\down}{\downarrow}

\begin{document}


\title{Quantum information processing using nuclear and \\ 
electron magnetic resonance: review and prospects}
\author{J. Baugh}
\author{J. Chamilliard}
\author{C. M. Chandrashekar}
\author{M. Ditty}
\author{A. Hubbard}
\author{R. Laflamme\footnote{email address : laflamme@iqc.ca}}
\author{M. Laforest}
\author{D. Maslov}
\author{O. Moussa}
\author{C. Negrevergne}
\author{M. Silva}
\author{S. Simmons}
\author{C. A. Ryan}

\affiliation{Institute for Quantum Computing, University of Waterloo, 
Waterloo, ON, N2L 3G1, Canada.}

\author{D. G. Cory}
\author{J. S. Hodges}
\author{C. Ramanathan}

\affiliation{Department of Nuclear Science and Engineering,
  Massachusetts Institute of Technology, 77 Massachusetts Ave,
  Cambridge, MA 02139}

\date{\today}
\begin{abstract}
  Quantum information is an exciting field promising a revolution in
  information processing.  A key ingredient for the advancement of
  this field is the development of technologies that can implement
  quantum information processing (QIP) tasks.  This paper describes
  recent progress using nuclear magnetic resonance (NMR) as the
  platform. The basic ideas of NMR quantum information processing are
  detailed, examining the successes and limitations of liquid and
  solid state experiments.  Finally, a future 
  direction for implementing quantum processors is suggested,
  utilizing both nuclear and electron spin degrees of freedom.
\end{abstract}

\maketitle

Quantum information processing is changing our fundamental
understanding of what information is and how it can be
manipulated. Recent work has lead to experimental proof-of-principle
demonstrations of the control of small quantum systems, and new
devices capable of harnessing larger quantum systems could change the
technological landscape of the 21st century. There exist many
proposals to physically realize such a quantum processor and presently
a few of these models are able to manipulate quantum bits (qubits):
quantum systems whose observables are given by the Pauli matrices.
A qubit encodes a fundamental unit of quantum information. One such
proposal uses the nuclear and electron magnetic moments and is the
subject of this short review.

Many atoms nuclei possess a magnetic moment. When placed in an
external magnetic field, these moments result in discrete energy level
systems that can be manipulated with resonant electromagnetic
radiation leading to NMR.  It was first observed more than half a
century ago by Purcell and Bloch \cite{PTP45a,Blo46a}. It has become a
powerful analytical tool with many applications such as
non-destructively determining molecular structures for chemistry
as well as static and dynamic imaging in both industry and medicine.

Many nuclei such as $^1$H or $^{13}$C are spin-1/2 quantum system.  
These are ideal qubits.  The initial
state of the system is obtained by allowing it to thermalize.  By
irradiating these nuclei at the appropriate frequency it is possible
to rotate their individual states one at a time leading to the so
called single-qubit gates. Nuclei affect each other through
interactions that can also be controlled, leading to two-qubit
gates. By composing these two sets of gates we can reach any unitary
transformation. This is known as universality \cite{KLM07a}.  After an
algorithm consisting of a series of these gates, the final state can
be observed by measuring the current induced by the rotating magnetic
moments of the sample in a conducting coil.  Algorithms are to be
performed on a timescale shorter than the characteristic decoherence
time of the system, which is about 100 times the gate time in liquid
state NMR.

The strength of NMR technology is the exquisite control that can be
implemented in multi-qubit systems.  Many decades of radio-frequency
(RF) engineering improvements today provide sufficient control
to implement experimental benchmarks such as quantum error correcting
codes and simulations of quantum physics systems, among many
others. In this paper, we first give a brief review of both liquid and
solid state NMR and then discuss recent work using both electron and
nuclear spins to form more powerful information processing
devices. The latter work provides a pathway to reach high purity
states which has been a weakness of the liquid state NMR proposal.

The rest of the paper is organized as follows:
sections~\ref{2a}--~\ref{2e} give a detailed introduction to the
concepts of NMR based quantum information processing in the liquid
state environment; sections~\ref{2f}--~\ref{2g} introduce more
advanced ideas of quantum control based on refocusing and optimal
control methods; section~\ref{3} describes the solid state
environment, a three-qubit solid state processor and experimental
results demonstrating high-fidelity control; section~\ref{4} discusses
a new direction involving systems of coupled electron and nuclear
spins, including motivations, prospects, and recent progress; and
finally section~\ref{5} concludes the paper.

\section{Liquid state NMR}

\subsection{Magnetic interactions}
\label{2a}

In the semi-classical picture, the spin of a nucleus behaves like the
dipolar moment of a magnet possessing angular momentum parallel to its
magnetic moment.  When placed in a constant magnetic field pointing along a
certain direction, (customarily defined as the $z$ direction) the
dipolar moment precesses around this axis.  The frequency of this
precession is called the Larmor frequency and is dependent on
the external magnetic field, the nuclear isotope and its chemical
environment within the molecule.  For quantum information purposes, we
are mainly interested in spin-1/2 nuclei (e.g. $^1$H, $^{13}$C,
$^{15}$N, $^{19}$F, $^{29}$Si and $^{31}$P to name a few).

Placed in magnetic fields generated by modern superconducting magnets, different
species of nuclei have differences in Larmor frequency on the order of
MHz. For example, the Larmor frequency of $^1$H is about 500 MHz in a
11.7 Tesla magnet, while that of $^{13}$C is about 125 MHz.  Depending
on the symmetry of the molecule, two nuclei of the same species can
either have the same Larmor frequency, or can have a frequency
difference (called chemical shift) ranging from a few Hz to
several kHz.  Typical liquid state NMR experiments involve an
ensemble of around $10^{20}$ identical molecules dissolved in a solvent whose 
effect on the nuclear magnetic moments of our molecules can be neglected.

When two spins are spatially close, their dipolar moments interact
with each other. The strength of this coupling is dependent
on the distance between the two spins and their relative orientation with
respect to the external magnetic field.  In a liquid, the molecules move and rotate around each other on a much shorter time
scale than the interactions occurring between them.  This causes the
intermolecular and intra-molecular dipolar interactions to average to
zero on the NMR time scale (i.e. the Larmor period time scale).  In solid state NMR, however, dipolar interactions remain and can be controlled as discussed in
section~\ref{3}.

Within the same molecule, there are still interactions between the
spins in the liquid state. If the wavefunctions of bonding electrons
overlap spatially with a pair of nuclear spins, the electron mediates
an effective interaction between the nuclear spins.  This interaction
is independent of the external magnetic field and the orientation of
the molecule, which inspires its name: scalar coupling (also called
indirect spin-spin coupling, or $J$-coupling).

\subsection{The NMR Hamiltonian}
\label{2b}

As mentioned above, in liquid state NMR the intermolecular spin
interactions are suppressed.  This causes the molecules to be
effectively isolated from each other, and therefore a description of
the spin dynamics of an ensemble of molecules is well approximated by
the spin dynamics of a single molecule.  If we consider a molecule
containing $N$ spin-1/2 nuclei in one of the molecules, then the
natural Hamiltonian of this system in a large homogeneous magnetic
field $\vec{B}_0$ pointing in the $z$ direction is given by
\be \label{eq:1}
\ham_{nat}&=&\frac{1}{2}\sum_{i=1}^{N}2\pi\nu_i^L {\sz^i}+
\frac{\pi}{2}\sum_{i<j}J_{ij} {\sz^i}
{\sz^j} 
\ee 
where $\nu_i^L=\omega_i^L/2\pi=\gamma_i |\vec{B}_0|$ is the
Larmor frequency of the $i^{th}$ nucleus with gyromagnetic ratio
$\gamma_i$, $J_{ij}$ is the coupling strength between nucleus $i$ and
$j$ and $ {\sz^i}$ is the $z$ Pauli matrix of the
$i^{\textrm{th}}$ spin.

The first term in the Hamiltonian describes the precession of the
spins due to their coupling to the external magnetic field, while the
second term describes the $J$-coupling between pairs of nuclei.  This
Hamiltonian corresponds to the weak coupling limit, where we assume
that the chemical shifts between coupled spins are much greater than
their respective couplings, i.e. $|\nu_i^L-\nu_j^L|>>J_{ij}/2$.  If this
approximation is not valid, we need to use the full coupling
Hamiltonian ${\sx^i} {\sx^j}+
{\sy^i} {\sy^j}+ {\sz^i}
{\sz^j}$ in place of ${\sz^i}
{\sz^j}$.  The exact values of the Hamiltonian parameters
are determined by fitting experimental data.

\subsection{Single-spin control}
\label{2c}

For quantum information processing, we need to be able to perform
arbitrary manipulations of a single spin, which is equivalent to
arbitrary rotations about any axis. As an example, consider the
application of a magnetic field $\vec{B}_1$ perpendicular to the $z$
axis which oscillates at the nuclear spin's Larmor frequency: 
\be
\B&=&|\B|\left(\cos(\omega^{rf}t)\,\vec{x}+\sin(\omega^{rf}t)\,\vec{y}\right)
\ee 
where $\omega^{rf}=2\pi\nu^{rf}$ is the angular frequency of the
field. In the rotating frame of the nucleus (i.e. the frame rotating
at the same frequency as the spin), $\vec{B}_1$ will appear as a
constant field pointing along its rotating $x$ axis. The spin will
start to precess about this axis.  Rotation about any axis in the
$xy$-plane is possible by adjusting the phase of the $\vec{B}_1$
field, e.g. $\omega^Lt \to \omega^Lt + \phi$, which will create a
rotation around the axis making an angle $\phi$ with the $x$ axis.  In
the laboratory, such a rotating field can be applied by sending a
radio-frequency (RF) pulse of a particular duration and phase to a
conducting coil surrounding the sample, calculated according to the
rotating wave approximation (see \cite{Lev01b} for more details).

To better understand this phenomenon from the viewpoint of quantum
mechanics, consider the rotating frame picture: suppose the spin is in
the state $\ket{\psi(t)}$, and define the state in the rotating frame
of the pulse with angular frequency $\omega^{rf}$ as
\be
\rpsi&=&\R_z(-\omega^{rf}t)\ket{\psi(t)}\nonumber\\
&=&\R_z(-\omega^{rf}t)e^{-\frac{it}{\hbar}\ham_{nat}}\ket{\psi(0)} \nonumber\\
&=&e^{\frac{i}{\hbar}\sz\frac{\omega^{rf}}{2} t}e^{-\frac{it}{\hbar}\ham_{nat}}\ket{\psi(0)}\label{rotstatedef}\\
&=&\ket{\psi(0)}, \,\,\textrm{for a single spin with
  $\omega^{rf}=\omega^L$}.
\ee 
If we apply a time derivative to
equation~\ref{rotstatedef}, it can be shown that the state in the rotating
frame $\rpsi$ evolves according to the Schr\"odinger equation with the
new Hamiltonian 
\be {\ham}_r&=&\R_z(-\omega^{rf}t)\ham_{nat}
\R_z(\omega^{rf}t)-\frac{\omega^{rf}}{2}\sz.
\ee 
When an RF pulse
with phase $\phi$ is applied to the spin, the laboratory frame
Hamiltonian is:  
\be \ham&=&\frac{\omega^L}{2}\sz +
\frac{\omega^{nut}}{2}
\left(\cos{(\omega^{rf}t+\phi)}\sx+\sin{(\omega^{rf}t+\phi)}\sy\right)
\ee 
where $\omega^{nut} = \pi\gamma_i | \vec{B}_1 |$. 
In the rotating frame this becomes 
\be
{\ham}_r&=&
\frac{1}{2}(\omega^L-\omega^{rf})\sz+
\frac{1}{2}\omega^{nut}(\cos\phi\,\sx+\sin\phi\,\sy).
\ee 
Thus, if the RF pulse is at the same
frequency as the spin, the spin will see a constant field in the
$xy$ plane, and will precess about it.  The rotation
angle $\theta$ is determine by the interval $\tau$ during which the RF
field is applied, according to $\theta=\omega^{nut}\tau$.

\subsection{Adding a second spin}
\label{2d}

It is also possible to independently control two spins with different
Larmor frequencies.  Applying an RF pulse at the frequency of the
first spin, the rotating frame Hamiltonian is given by 
\be
\label{eq:8}
\widetilde\ham_{nat}&=& \frac{1}{2}\omega_1^{nut}\sx^1+
\frac{1}{2}\omega_2^{nut}\sx^2+
\frac{1}{2}(\omega_2^L-\omega_1^L)\sz^2+ \frac{\pi}{2}J_{12}\sz^1\sz^2
\ee where we have set $\phi=0$ for simplicity. While the first spin
undergoes a rotation around the $x$ axis, the second spin experiences
a field with an additional non-zero $z$ component.  This is called the
off resonance effect. If we consider the case where
$\omega_2^L-\omega_2^L>>\omega_1^{nut}$ then the second spin rotation
around the $x$ axis will average to zero during the time the first
spin has completed its rotation. Typically, $\omega^{nut}$ is smaller
than 1 MHz, so this condition is automatically satisfied if the two
nuclei belong to different species.  If the spins are of the same
species, this condition can also be satisfied if a very low amplitude
pulse is used due to the small nutation frequency. In this case, one
drawback is that the pulse will necessarily take much longer to
achieve the same angle of rotation, and if the two spins have a
significant coupling constant $J_{12}$ coupling effects might
introduce significant errors and therefore limit our control.

Fortunately, there exist well known techniques to address different
nuclei of the same species with high precision.  The most common
technique is to control the spins using shaped pulses.  The frequency
response to the pulse will depend on the pulse shape (Fourier theorem)
and so by applying the pulse with a time varying power we can control
the power spectrum of the pulse.  For example, if a Gaussian shaped
pulse is applied at frequency $\omega^{rf}$, then only spins within a
Gaussian distribution of frequencies around $\omega^{rf}$ will respond
to this RF field. Therefore, if the height and the length of the
Gaussian pulse is carefully chosen, one spin can be ``addressed'',
causing negligible effects to others. This technique permits control
of spin pairs with smaller chemical shift differences in shorter
periods of time, hence allowing stronger coupling. The length of a
Gaussian pulse is proportional to the inverse of the chemical shift
between the spins.  Therefore, in the limiting case of small chemical
shift differences and large $J$-coupling values, control of the qubits
is more difficult.

For most liquid state experiments on a few spins, where chemical
shifts are comparatively large and $J$-couplings are small, the use of
Gaussian pulses is sufficient to achieve very high precision spin
rotations.  The situation becomes more complicated when there are more
homonuclear spins (implying smaller chemical shift differences on
average), or stronger coupling like in solid state or liquid
crystal
environments.  It is still possible to
overcome these drawbacks by considering more complicated pulse shapes
and phase modulation. For example, in section~\ref{2g}, we will
describe how it is possible to find shaped pulses that can implement
any desired evolution by simulating the full quantum dynamics.

\subsection{The controlled-NOT operation}
\label{2e}

\begin{figure}
\scalebox{0.4}{\includegraphics{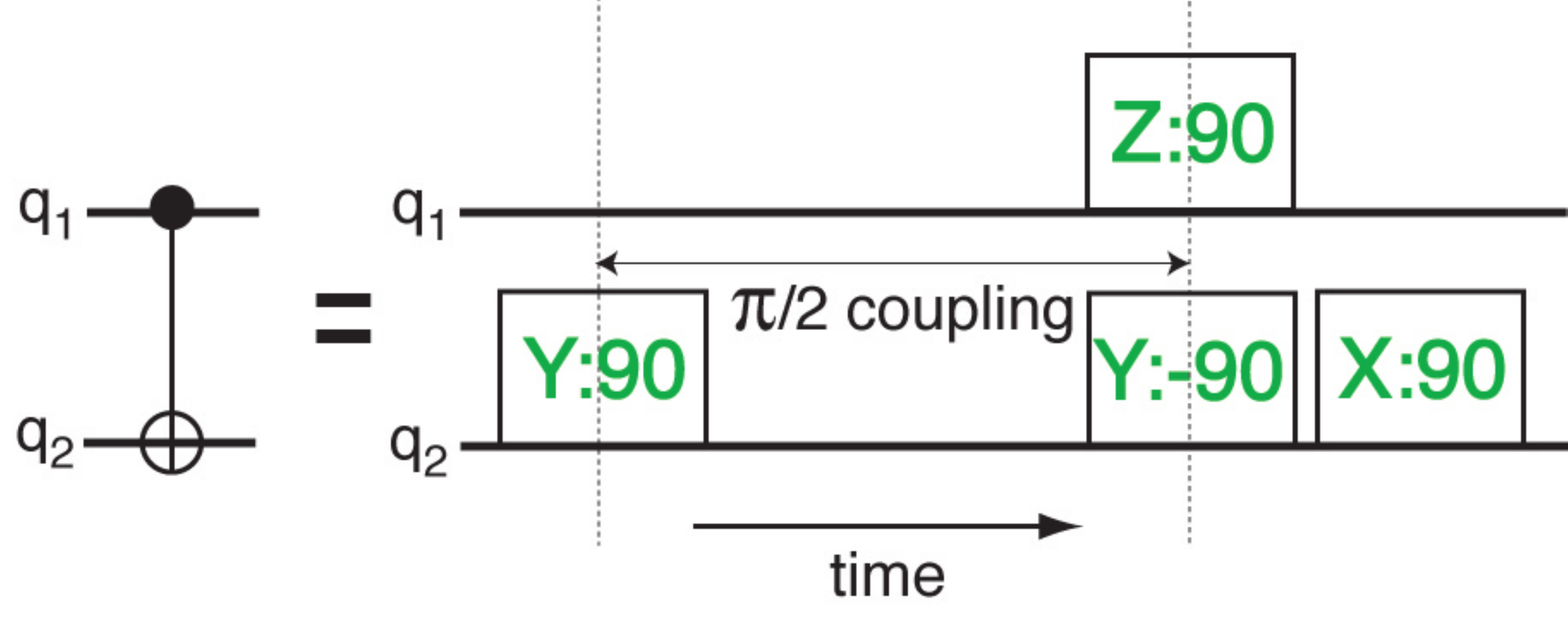}}
\caption{Implementation of a controlled-NOT gate in liquid state NMR. 
The left circuit is the quantum circuit representation of a controlled-NOT 
gate with control qubit q$_1$ and target qubit q$_2$.  On the right is the 
NMR implementation of such a  gate by combining single qubit rotations and 
the natural two spin interaction of the system. The single qubit rotation
 properties are given in the rectangles, e.g. $X:90$ is a short notation 
for $R_x(\frac{\pi}{2})=e^{-i\frac{\pi}{4}\sigma_x}$.  Notice that the size 
of the rectangles are not to scale for liquid state NMR, where the RF 
pulses are usually much shorter than the time to implement a 
$\frac{\pi}{2}$ J-coupling
\label{fig:cnotliquid}}
\end{figure}

In the previous subsection we discussed a method used to independently control
different spins. In order to perform quantum computing, we need to achieve 
universal control and hence be able to have spins interact with each other. 
A two-qubit gate that is useful for quantum information processing is 
the controlled-NOT, which acts as
\be
\ket{00}\to \ket{00} &,&\ket{10}\to\ket{11}\nonumber\\
\ket{01}\to\ket{01} &,&\ket{11} \to \ket{10}.
\ee
The operation must flip the target qubit (second bit) if and
only if the first qubit is in the state $\ket{1}$. In NMR, $\ket{0}$
and $\ket{1}$ are associated with the state of the spin pointing up,
$\ket{\up}$ or pointing down $\ket{\down}$ respectively.  If we look at
the Hamiltonian in equation~(\ref{eq:1}), and consider its 
effect on spin 2 depending on whether the state of spin 1 is up or
down, we obtain an effective Hamiltonian for the second spin:
\be
\ham_{\up}^{2}=\frac{1}{2}(\omega_2^L+\pi J_{12})\sz^2 \\
\ham_{\down}^{2}=\frac{1}{2}(\omega_2^L-\pi J_{12})\sz^2.
\ee
Therefore, spin 2 will rotate slower or faster depending
on the state of spin 1.  If the coupling evolves for a time 
$\tau=\frac{1}{2J_{12}}$, 
we obtain a controlled-Z rotation of $\frac{\pi}{2}$ degrees, 
which can be transformed 
into a controlled-NOT by a few single spin pulses applied before and after (see 
figure~\ref{fig:cnotliquid} for the complete sequence to implement a 
controlled-NOT).

In the previous sections, we demonstrated that it is possible to
implement any rotation around an arbitrary axis in the $xy$-plane, as well
as perform a controlled-NOT gate with two spins.  These two conditions
are sufficient to perform universal quantum computing.  For practical purposes, 
it is convenient to use only $\frac{\displaystyle\pi}{\displaystyle
  2}$ and $\pi$ RF pulses as they are easier to calibrate.  Arbitrary angle 
rotations can still be applied because $z$-axis rotations can be obtained by 
changing the definition of the rotating frame, which is equivalent to 
changing the phase of subsequent pulses.  It can be verified that 
\be \label{phasetrack}
R_x(\frac{\pi}{2})R_z(\theta)=R_z(\theta)R_{{n}}(\frac{\pi}{2})
\ee
where the vector
${n}=\cos{\theta}\,{x}-\sin{\theta}\,{y}$.  Therefore, since $z$
rotations commute with the internal Hamiltonian of the system, we may
commute all the $z$ rotations to the end of the pulse sequence and
compensate for any remaining $z$ rotation during the post-processing
of the data.  Moreover, the overall $z$-rotation takes no time and is
far more precise when using this procedure because the RF phase has a
higher accuracy than the RF amplitude in modern NMR spectrometers.

\subsection{Refocusing and control techniques}
\label{2f}

\begin{figure}
\scalebox{0.4}{\includegraphics{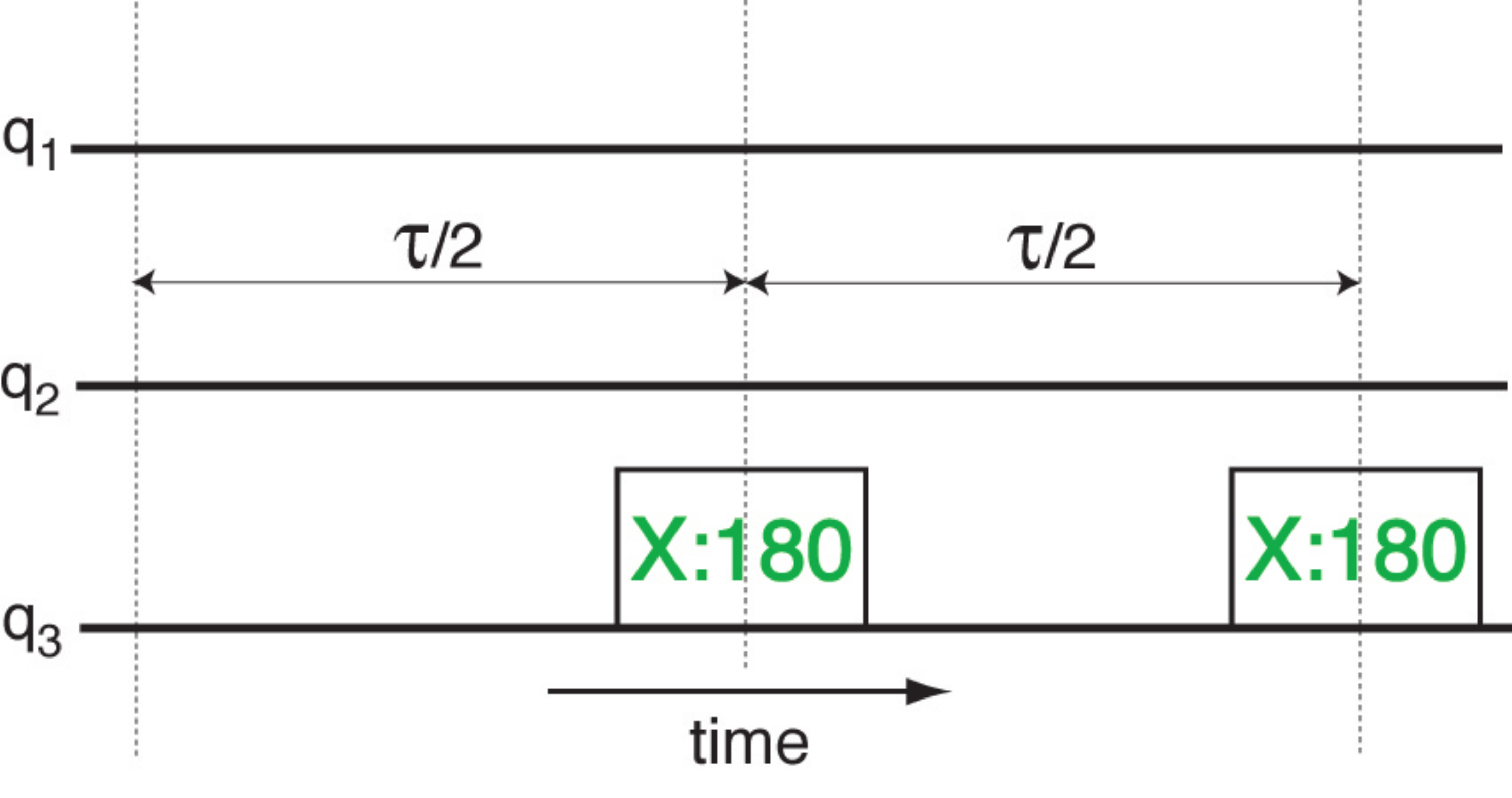}}
\caption{In NMR, the coupling between spins is always active.  It is
  possible to refocus 2-qubit interactions using special pulse
  sequences. An example is given above, where halfway through a period
  $\tau$, a $\pi$ pulse on one of the nuclei is implemented that
  reverses the direction of its spin. Note that when we have a
  coupling of the form $\hat{\sigma_z}^{(1)} \hat{\sigma_z}^{(2)}$ the
  effect of the pulse can be thought of as reversing the sign of the
  coupling, and thus allows to cancel the interaction that occurred
  during the first $\tau/2$ period. This pulse sequence effectively
decouples the third qubit from the system. while leaving the coupling
between q$_1$ and q$_2$ unchanged.  This situation is mathematically
treated in equation~(\ref{eq:refocus})
\label{fig:refocus}}
\end{figure}

In an NMR system, spins constantly couple to each other, and we must
``turn off'' these couplings on demand to implement generic quantum
gates.  For example, consider a three spin system in which we wish to
implement a controlled-NOT between the first and second qubits.  As
mentioned above, a $\frac{\pi}{2}$ $J$-coupling between spins 1 and 2
is needed, which is accomplished by allowing the system to evolve
under the natural Hamiltonian for a time
$\tau=\frac{1}{2J_{12}}$. During this time spin 3 will also couple to
spins 1 and 2, giving an unwanted evolution.  However, if we apply a
$\pi$ pulse on the third spin half-way through the free evolution (at
time $\frac{\tau}{2}$), this spin will effectively decouple from the
other two spins and, upon an extra $\pi$ pulse at the end of the
evolution (at time $\tau$), it will be returned to its initial state
(this pulse sequence is shown in Figure~\ref{fig:refocus}).
Considering only the interaqction term of the 
intramolecular component of the Hamiltonian, we
can write the evolution of the system as:
\be
\U(t)&=&R^{\dagger 3}_x(\pi)
       e^{-i\ham\tau/2}R^{3}_x(\pi)e^{-i\ham\tau/2}\nonumber \\
&=&(i\sx^{3})
e^{-i\frac{\pi\tau}{4}(J_{12}{\sz^1}
{\sz^2}+J_{13}{\sz^1}{\sz^3}+J_{23}{\sz^2}{\sz^3}} (-i\sx^{3})
e^{-i\frac{\pi\tau}{4}(J_{12}{\sz^1}{\sz^2}+J_{13}{\sz^1}{\sz^3}+J_{23}{\sz^2}{\sz^3})}
\nonumber\\
&=& 
e^{-i\frac{\tau}{2}
  (J_{12}{\sz^1}{\sz^2}-J_{13}{\sz^1}{\sz^3}-J_{23}{\sz^2}{\sz^3})} 
e^{-i\frac{\tau}{2}
  (J_{12}{\sz^1}{\sz^2}+J_{13}{\sz^1}{\sz^3}+J_{23}{\sz^2}{\sz^3})}\nonumber\\
&=& e^{-i\frac{\pi\tau}{2} J_{12}{\sz^1}{\sz^2}},
\label{eq:refocus}
\ee
where $R^{3}_x(\pi)$ is the operator of a $\pi$-pulse about the
$x$-axis on spin 3.  This is called a refocusing scheme and can be
readily generalized to any number of coupled spins, i.e. a $\pi$ pulse
on spin $i$ will effectively decouple it from all the other spins.
This scheme can also be efficiently generalized to decouple $m$ spins
from each other and from the $N-m$ remaining spins. In practice, the
situation is more complex.  For example, one must keep track of the
Zeeman evolution of all the spins (which is called phase
tracking). This evolution can be taken into account by changing the
phase of subsequent pulses according to the relation given in
equation~\ref{phasetrack}.

For systems involving up to a few spins, pulse phases and decoupling
sequences are derived by hand, but for molecules containing
greater number of spins, these calculations become tedious 
and computer assisted techniques are used~\cite{BJK+05a}. Efficient
classical algorithms can be implemented that optimize pulse sequences
with respect to phase and residual coupling errors.

\begin{figure}
\scalebox{0.4}{\includegraphics{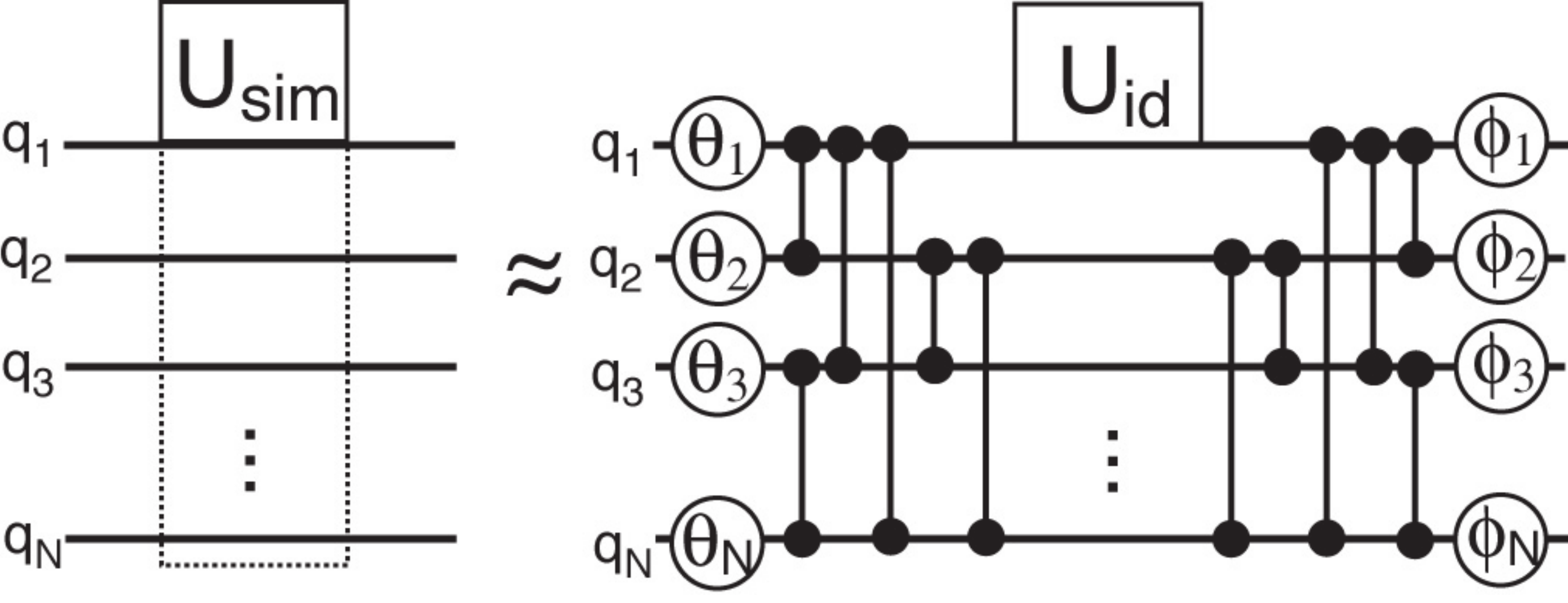}}
\caption{A selective pulse designed to implement a single-qubit rotation in 
an $N$-spin system will, in general, also affect the other spins.  This can 
be studied in small systems by simulating the full quantum dynamics to 
obtain the unitary $U_{sim}$. The unwanted evolution of the non-target 
qubits is represented by the broken line on the left figure.  If the pulse 
is carefully designed so that its implementation is very close to the ideal 
desired unitary $U_{id}$, the error can be efficiently estimated to first 
order by phase errors (represented by $\theta_i$ and $\phi_i$) and 
coupling errors occurring before and after the pulse. 
\label{fig:approx}}
\end{figure}

A major source of pulse errors are off
resonance and coupling errors.  It is possible to estimate and
compensate for these errors by evaluating the first-order coupling and
phase errors generated by a pulse.  This is done by assuming that the
real pulse can be decomposed as the ideal pulse preceded and followed
by phase and coupling errors (see figure~\ref{fig:approx}).  Since the
error terms all commute with each other, they can be estimated using
pairwise spin simulations, which requires reasonable computational
resources, i.e. is efficient as we scale the number of nuclei.  
With small $J$-couplings and short pulses it is reasonable
to expect error rates below a fraction of a percent for each pulse.

Once the errors generated by each pulse are known they can be taken
into account and corrected for by optimizing the durations of the free
evolution periods and the timing of the refocusing pulses.  Such an
algorithm can also perform phase tracking and modify the pulse phases
accordingly. Very high gate fidelities have been demonstrated using
this efficient pairwise simulation technique.

\subsection{Optimal control for strongly coupled spins}
\label{2g}

In some cases, spins are so strongly coupled that the approximation of
${\sz^i}{\sz^j}$ couplings used above breaks down.
In those cases another technique can be used: strongly modulating
pulses designed using numerical optimal control
techniques~\cite{FPB+02a,KRK+05a}.  For systems containing about less than
ten qubits, we can find extremely high fidelity and robust control 
by applying optimal control principles.  Just as classical optimal
control theory can tell how to best steer a rocket, quantum optimal
control gives the tools to best steer a quantum system to a desired
unitary gate.  Quantum optimal control has been used for some time in
the context of driving chemical reactions with shaped laser pulses
\cite{RVM+00a}.  There, the goal is to maximize the
transfer from a known initial state to a known final state.  In the
context of quantum computing, we do not necessarily know what the
input state will be, and so we must find unitary gates which will work
correctly for any input state.

\begin{figure}
\scalebox{0.8}{\includegraphics{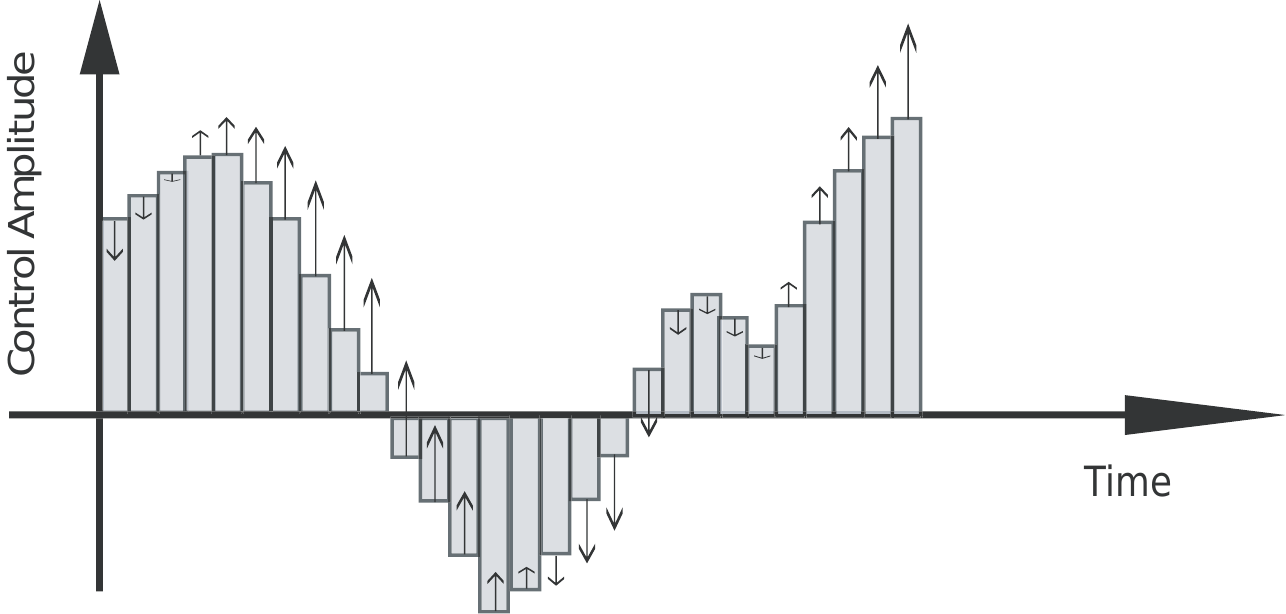}}
\caption{An example section of a
GRAPE pulse.  The bars show the control amplitude which is constant
for each time step. The arrows show the derivatives from the GRAPE
algorithm at each point which tell us how to update the pulse for the
next iteration (adapted from \cite{KRK+05a}).}
\label{fig:grape} 
\end{figure}

The Hamiltonian at any point in time can be written down as
\begin{equation}
\ham_{tot}(t) = \ham_{nat} + \ham_{rf}(t),
\end{equation}
where, $\ham_{nat}$ is the time-independent, natural (or drift) Hamiltonian, and
$\ham_{rf}(t)$ represents the controllable time-dependent RF fields
which we can use to drive the evolution of the system.  The task is to
find the sequence of control fields that will produce the correct
unitary evolution.  For reasonably complicated systems analytical
solutions are beyond reach; however, using a numerical search we can
find control sequences with fidelities as high as $0.999999$.
The control fields are discretized at a suitable rate (see
figure~\ref{fig:grape}) and a random guess for the control sequence is
chosen.  We can simulate the evolution of the system under this
sequence and obtain the simulated unitary $U_{sim}$.  We can then
compare this to our goal unitary $U_{goal}$ through a fitness
function,
\begin{equation}
\Phi = \left|Tr\left(U_{sim}^\dagger U_{goal}\right)\right|^2
\end{equation}
which is equivalent to the state overlap fidelity averaged over all
input states \cite{FPB+02a}.  We can then use any optimization routine to search for
the highest $\Phi$.  

The first step in this direction for NMR
quantum information processing was the scheme of
Fortunato {\it et al.} \cite{FPB+02a} who, by limiting the form of the control
fields to a small number of constant amplitude, phase and frequency 
periods, were able to find high fidelity
control sequences through a simplex search.  Many more control periods
can be considered and the numerical search substantially sped up
through the use of standard optimal control techniques to obtain
information about the gradient of the fitness function at each point.
This is the GRadient Ascent Pulse Engineering or GRAPE algorithm
introduced by Khaneja {\it et al.} \cite{KRK+05a}.  From the simulation
information we can calculate approximate gradients of the fitness
function with respect to the control amplitude at each timestep (see
Figure~\ref{fig:grape}).  With this gradient information we can update
the control fields by moving along the steepest ascent direction and
then repeat the procedure.  Convergence to a global maximum is of
course not guaranteed as this hill climbing algorithm is a local
search.  However, empirically we find we can achieve sufficiently high
fidelities from such local maxima.  Convergence of the algorithm
can be further improved by using non-linear conjugate gradients \cite{BH75a}.

\begin{figure}
\scalebox{0.35}{\includegraphics{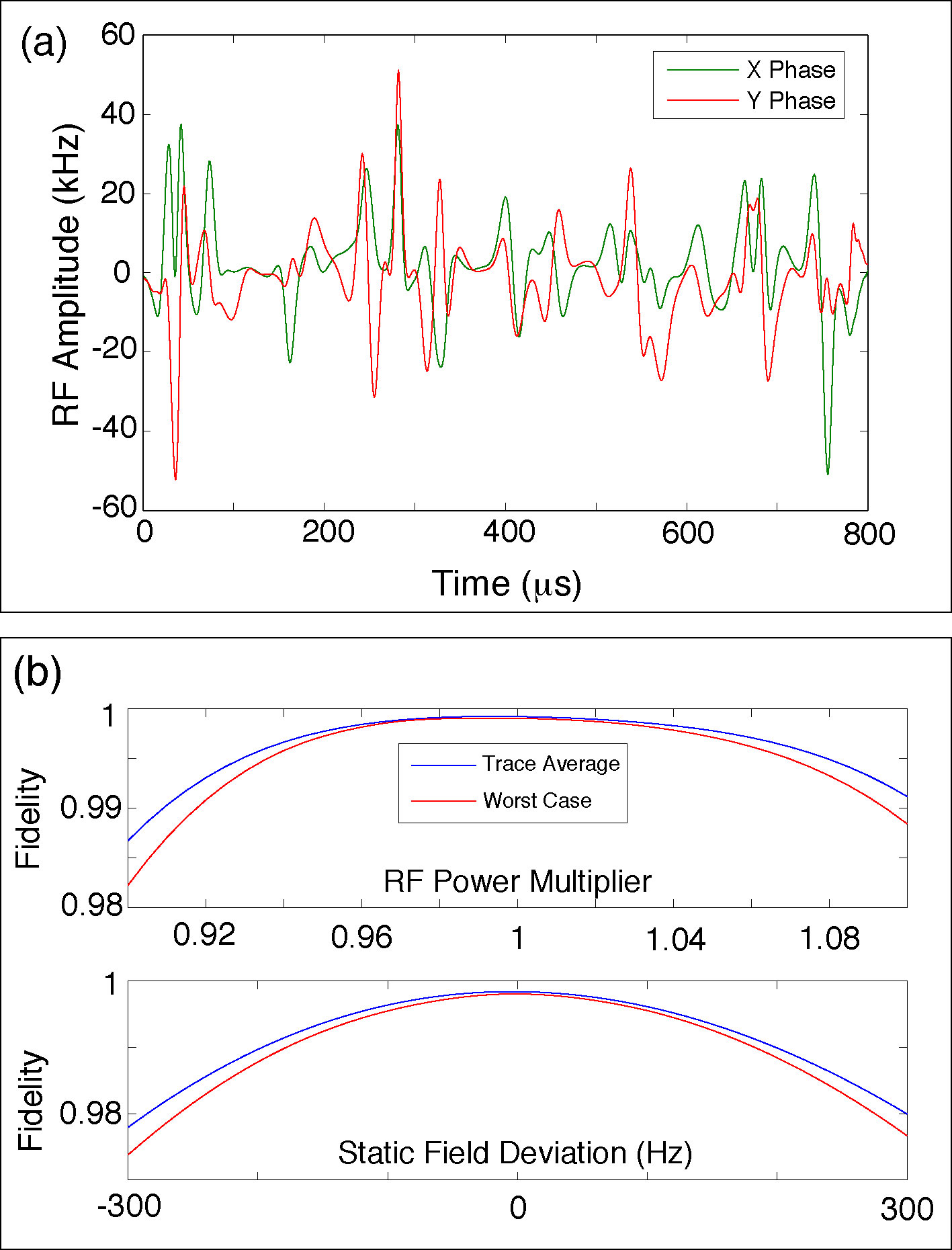}}
\caption{(a) An example of a GRAPE pulse designed to implement a
  controlled-NOT-NOT on the three qubit malonic acid processor. (b)
  The pulse is designed to be robust to inhomogeneities in RF
  amplitude and external magnetic field; the fidelity of the pulse is
  plotted versus both types of inhomogeneity.  The blue curve shows
  the fidelity of the gate averaged over all input states while the
  red curve shows the fidelity for the worst-case input state.  Note
  that the robustness of the gate fidelity to static field deviation
  is equivalent to self-refocusing of $T^{*}_2$ dephasing by the
  pulse.}
\label{fig:cnotnotpulse}
\end{figure}

The control sequences drive the system through a complicated and
non-intuitive path and small errors in our modeling of the system's
Hamiltonian might lead to a low fidelity pulse.  Fortunately, the
control sequences can also be made robust against
static inhomogeneities or uncertainties such as field inhomogenties
and amplitude miscalibration of pulses.
Robustness to both these effects for a particular
pulse is shown in figure~\ref{fig:cnotnotpulse}b.

Recent work on a 12-qubit liquid state system that compared this
method with the more traditional approach outlined in previous
sections can be found in \cite{NMR+06a}.  Not surprisingly, the
control using strongly modulating pulses was more precise but at an
exponential cost in searching for the pulse sequences.

The scheme described above can be generalized if we can break the
molecule into strongly coupled subsystems which are themselves weakly
coupled, and thus it is possible to perform the decomposition described
in previous sections between subsystems using pairwise simulation of
subsystems. Gates to implement rotations within the subsystem will
however require  numerical optimization methods such as GRAPE.  The details of
this hybrid method are presently under investigation.

The ability to control liquid state nuclear spin systems has allowed
implementation of a variety of benchmarking experiments and algorithms
on small qubit registers.  For example, we have developed sufficient
control to implement quantum error correcting protocols
\cite{CPM+98a,LSB+07a,CSH+00a,KLM+01a,BVF+04a}, the simulation of quantum
systems \cite{CYC04a,NSO+04a,HSV+01a,TSS+99a,STH+99a} and the benchmark 12 qubit quantum
processor \cite{NMR+06a}. An extensive list of these and other
experiments can be found at http://arxiv.org/.

\section{Solid State NMR}
\label{3}

Solid state NMR presents a very different spin environment in
comparison to liquid state NMR.  
Dipolar couplings do not average to
zero in solids and in fact tend to dominate the NMR spectra, usually
through significant broadening of the resonance peaks.  As an
illustration, consider an array of identical spins on a crystalline
lattice: any one nuclear spin couples most strongly to its nearest
neighbors, then to next-nearest neighbors, and so on (the dipolar
coupling is long range, falling off as $r^{-3}$ where $r$ is the
distance between spins).  At low spin polarizations, each coupling
induces a splitting of the spin's resonance energy into two components
of slightly higher and lower energies.  Averaging over the ensemble of
spins, a smoothly broadened resonance is observed, whose width is
comparable to the size of the nearest-neighbor coupling.  These
couplings can be quite large compared to the $J$-couplings observed in
liquids; for example, a proton pair separated by 2\AA \hspace{0.1mm}
has a maximal dipolar coupling frequency of 15 kHz, whereas
proton-proton $J$-couplings are usually 200 Hz or less.

The solid state spin environment would at first appear to be less than ideal
for quantum information processing, as the strong spin-spin
interactions would lead to rapid effective loss of coherent quantum information. 
However, this situation can be remedied in two
ways: (1) the evolution due to time-independent interactions is, in
principle, completely refocusable by appropriate RF pulse sequences and
(2) the spatial dilution of spins in the sample can be used to minimize
spin-spin interactions.  For the three-qubit processor described in section~\ref{3b}  based
on the malonic acid single crystal, we take advantage of both spin
dilution and NMR refocusing techniques to maintain good qubit
coherence. Additionally, spin-lattice relaxation times $T_1$ can be
very long (minutes to hours) in solids, particularly for spin-1/2
nuclei in crystalline solids at low temperatures and with low impurity or defect densities. 
We generally have $T_1>T_2 >> T^*_2$, where
$T^*_2$ is the dephasing time in the absence of refocusing, and $T_2$
is the ``intrinsic'' coherence time which is usually determined in practice for solid state 
systems by the quality of the refocusing controls. With perfect control, we should obtain $T_2 =2T_1$, which
would allow for many thousands of quantum gates to be performed within the characteristic
coherence time scale.  Of course, in practice, control errors and other experimental
imperfections greatly limit the effective spin coherence time.

Besides the potential for much longer coherence times, solid state NMR
also offers the prospect of greatly increasing spin polarization by
cooling the sample to low temperatures. Nuclear Zeeman energies at
attainable magnetic fields are, however, on the order of
1 $\mu$eV or less, and such small energies require a temperature less than 
10 mK to reach full polarization by thermal equilibration.
Fortunately, nuclei may also be polarized dynamically at much more
accessible temperatures $\sim$1.5K, by driving polarization exchange
between nuclear spins and a small concentration of unpaired electrons
that have been introduced into the sample \cite{AG82a}.  Since the magnetic 
moment of the electron is about 660 times larger than that of the proton,
electron spins can be fully polarized in thermal equilibrium at such
temperatures.

More generally, nuclear and electron spins incorporated into a
solid lattice are thought to be promising elements for building
scalable quantum information processors \cite{Kan98a,JTS+07a,MM06a}.  Nanofabrication technologies are continually growing in precision and
sophistication, allowing ever more controllable nanoscale
devices. Nuclear spins can provide stable qubit storage,  and the
transfer of information to electron spins can allow for fast
operations and coupling to optical photons can provide initialization, readout
and long-range entanglement \cite{JTS+07a,JGP+04a,JGP+04b,CDT+06a}.  Controlling the nuclear and
electron spins with high fidelity using magnetic resonance should play
an important role in these schemes.

\begin{figure}
\scalebox{0.25}{\includegraphics{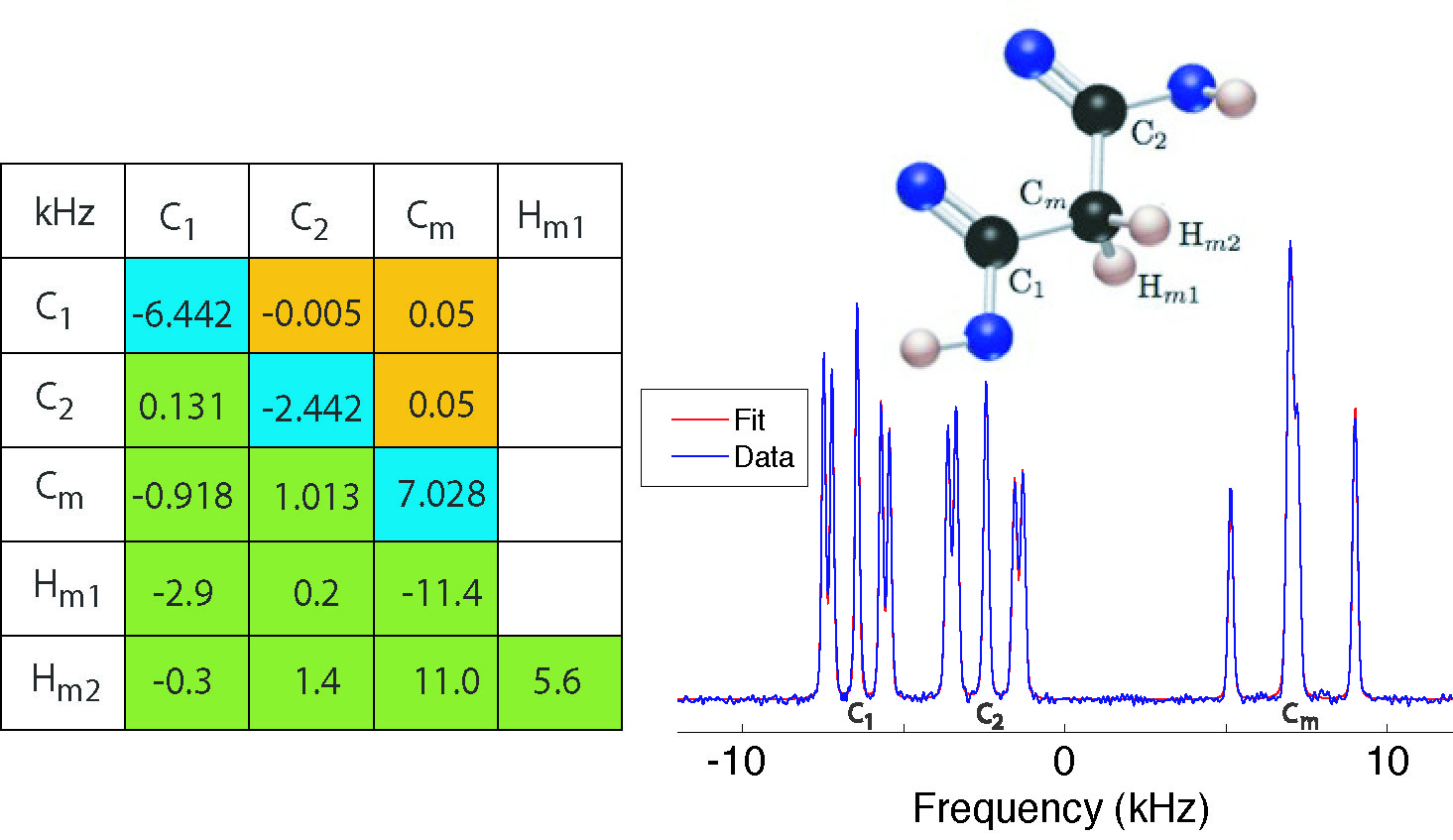}}
\caption{(right) Fitted spectrum of a malonic acid single crystal
  ($3.5\%$ $^{13}C_3H_4O_4$ concentration) at a particular orientation
  with respect to the external field. (left) Table showing the
  Hamiltonian parameters extracted from the spectral fitting: chemical
  shifts are along the diagonal, dipolar couplings in the lower
  off-diagonal (green), and $J$-couplings in the upper off-diagonal
  (yellow).  The dipolar parameters involving the methylene proton
  spins $H_{m1,2}$ have been estimated indirectly by using the
  $^{13}C$ parameters to precisely determine the crystal orientation,
  and knowledge of the crystal structure from neutron scattering data
  \cite{MK93a} to then calculate the couplings.}
\label{fig:malspect_hq} 
\end{figure}

\subsection{Three qubit solid state quantum information processor}
\label{3b}

The following example of a small quantum information processor is based on malonic acid,
$C_3H_4O_4$ \cite{BMR+06a}. The sample is a single crystal (grown
from aqueous solution) consisting of a mixture of $^{12}C_3H_4O_4$ and
$^{13}C_3H_4O_4$ (labeled) molecules; $^{13}C$ is spin-1/2 whereas
$^{12}C$ is spinless. The fraction of labeled molecules is kept to
$<10\%$ of the total, so that the $^{13}C_3$ molecules can function as
an ensemble of 3-qubit registers with relatively weak spin
interactions between registers.  It is important to note that all
molecules contain 4 spin-1/2 $^{1}H$ atoms, so that generally, a
decoupling RF sequence must be applied to the $^{1}H$ spins to isolate
the $^{13}C$ qubits from the $^{1}H$ spin
system. Figure~\ref{fig:malspect} shows a $^{13}C$ NMR spectrum (with
$^{1}H$ decoupling) obtained at an external field magnitude and
orientation at which the three carbons are well separated in Larmor
frequency (i.e. separately addressable) and the carbon-carbon
intramolecular dipolar couplings are relatively large, allowing for
fast coupling gates.  The spin Hamiltonian of the 3-qubit system is
determined directly by fitting a simulated spectrum to the measured
one.  The height differences between the absorption peaks are due to
the ``strong coupling'' effect, in which non-diagonal terms in the
coupling Hamiltonian (such as
$d_{12}(\sigma^1_x\sigma^2_x+\sigma^1_y\sigma^2_y)$) have frequencies
$d_{12}$ comparable to the difference in chemical shifts of the
coupled spins ($\Delta\omega^L(\sigma^1_z-\sigma^2_z)$).

When $d_{12}\sim \Delta\omega^L$ as in the present solid state system,
the $\sigma^i_z$ eigenstates (i.e. the computational
basis states $\ket{000}$, $\ket{001}$, etc.), are no longer the energy
eigenstates and the spin dynamics become more complex.  This poses a
challenge for universal quantum control: the scalable method described
in section \ref{2f} for generating quantum gate pulse sequences in
liquid state NMR is valid only when such mixing is negligible.
However, numerically optimized strongly modulating pulses can succeed
in providing sufficient control.

This control can be made robust with respect to sources of ensemble
incoherence. For example, the linewidths ($(T^{*}_2)^{-1}$) of the
resonance peaks visible in Figure~\ref{fig:malspect} are about two
orders of magnitude larger than those seen in a liquid
environment. This incoherence is mainly due to a combination of
residual dipolar couplings between $^{13}C_3H_4O_4$ molecules and
inhomogeneity of the external magnetic field over the sample
\footnote{This inhomogeneity is mainly caused by susceptibility
  mismatch between the crystal and the air/glass surrounding it, and
  the fact that it is non-spherical.}. Strongly modulating pulses can
be made sufficiently robust to these effects with minimal added
computational overhead, particularly when using the GRAPE algorithm
(see figure~\ref{fig:cnotnotpulse}).

\begin{figure}
\scalebox{0.5}{\includegraphics{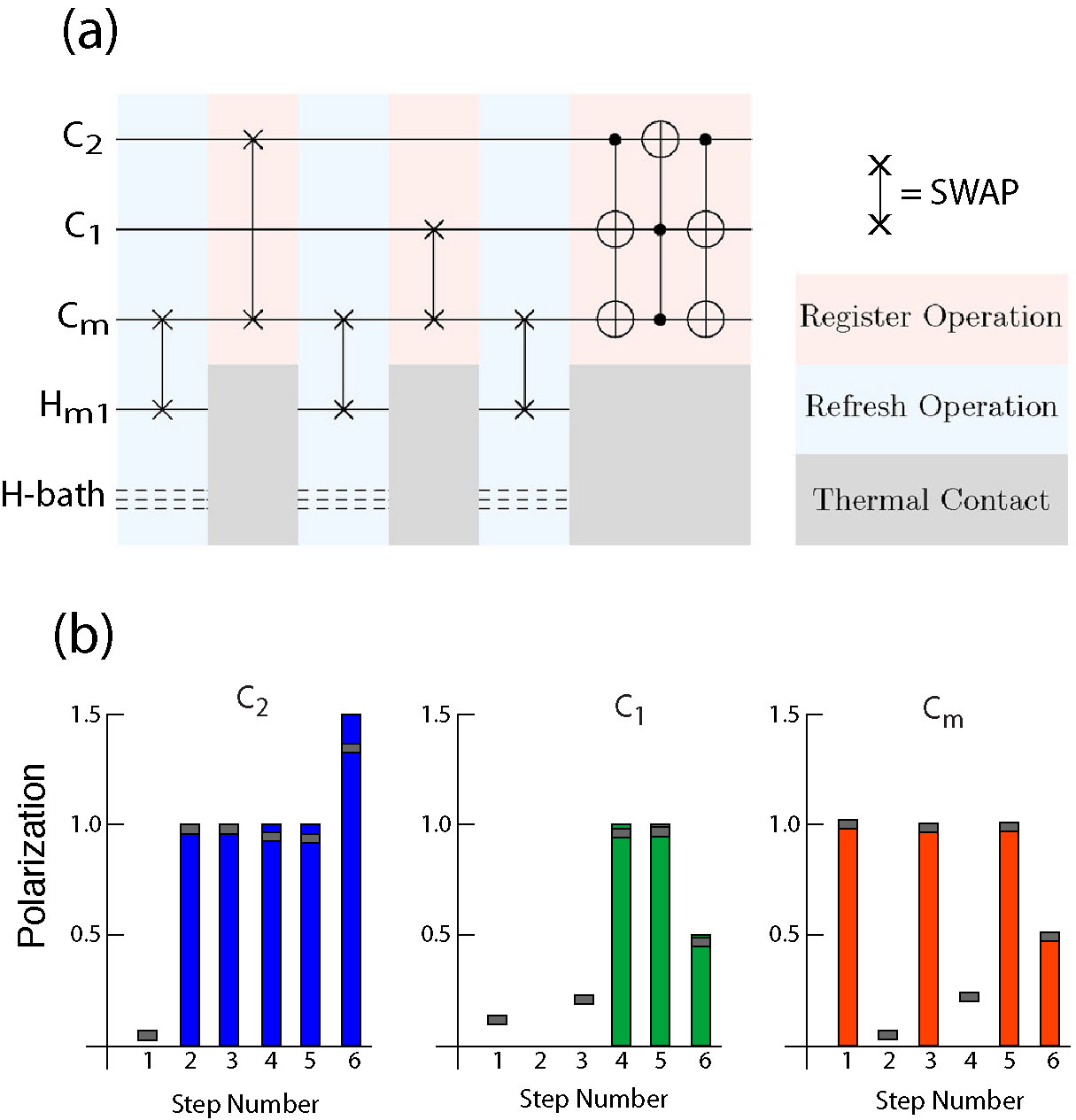}}
\caption{(a) Schematic quantum circuit diagram of the HBAC protocol
  implemented in malonic acid. Time flows from left to right.  The
  last step corresponds to a three-bit compression unitary operator, and is here decomposed into two controlled-NOT-NOT
  gates sandwiching a Toffoli gate \cite{Tof80a}. (b) Experimental results,
  in terms of measured qubit polarizations at each step. The full bars
  indicate the ideal values of the protocol, and the shaded bands give
  the experimental values (uncertainties are given by the widths of
  these bands).}
\label{fig:ac_results}
\end{figure}

\subsection{Algorithmic Cooling}
\label{3c}

As a computational demonstration of the malonic acid quantum
processor, we now consider the experimental implementation of a
heat-bath algorithmic cooling (HBAC) protocol \cite{BMR+05a}.  The aim
of the experiment is to amplify the polarization on one of the three
qubits, using both qubit-qubit operations and a controlled coupling to
a surrounding ``heat-bath'' \cite{SMW05a,BMR+02a}.  Note that HBAC is
a classical algorithm, since each step results in a classical state of
the system, however we use quantum gates in its implementation.
Strongly modulating pulses obtained using GRAPE are used to perform
gate operations on the qubits. The HBAC algorithm operates on the
three malonic acid qubits as well as the large surrounding $^{1}H$
spin system which acts as a spin-bath with a large effective heat
capacity.  Figure~\ref{fig:ac_results}a shows a schematic quantum
circuit diagram of the first six steps of the HBAC protocol
implemented in the malonic acid system.  The swap operations between
$C_m$ and $H_{m1}$ (labeled `refresh' in figure~\ref{fig:ac_results}a)
can be carried out either using a dual $^{13}C$/$^{1}H$ multiple-pulse
refocusing sequence \cite{BMR+05a,WGP82a} or by using Hartmann-Hahn
cross-polarization \cite{HH62a} with short contact time \footnote{The
  former approximates a universal quantum swap operation, whereas the
  latter is sufficient for swapping polarization.}.
  
The multi-qubit operations are carried out using strongly modulating
pulses.  During these operations, a strong continuous wave decoupling
field is applied to the $^{1}H$ spins, serving both to decouple them
from the carbons and to ``lock'' their collective magnetization along
a particular direction in the rotating frame (the latter is called
`spin-locking' and is used to prevent the magnetization from
dephasing).  This continuous wave field permits the protons to
interact with each other via the $^{1}H-^{1}H$ dipolar couplings. Such
couplings diffuse spin polarization \cite{ZC98a} through the
Hamiltonian terms of the form $\sigma^{1}_+\sigma^{2}_- +
\sigma^{1}_-\sigma^{2}_+$ \footnote{These terms involve the ladder
  operators defined as $\sigma_{\pm}=\sx \pm i\sy$}. Therefore, the
$H_{m1}$ polarization of the spins restore to the equilibrium value of
the $^{1}H$ bath after sufficient interaction time. The timescale for
such equilibration in this system is $\sim 50\times T^{*}_2 (H)
\approx 5$ms $<< T_1(H)$.

Experiments were performed at room temperature in the regime of low
nuclear spin polarization, and the aim was to demonstrate quantum
control similar to what has been achieved using liquid state NMR
quantum information processors.  The measured polarizations on each of
the qubit spins at each step of the algorithm are shown in
Figure~\ref{fig:ac_results}. The full bars indicate ideal values and
the shaded bands give the experimental values.  A rough indication of
the fidelity of the experimental protocol can be gauged by the ratio
of the final polarization of the target qubit $C_2$ to its ideal
value, $F\approx 1.39/1.5=92.7\%$. This yields an average error (here
loss of polarization) per step of about $1.5 \%$.  Given that the
inhomogeneous line-broadening effects (i.e. the natural incoherence of
the system) in solid state NMR is two orders of magnitude larger than
that in liquid state, this is an impressive degree of control.
Furthermore, these results indicate that such sequences applied to
single quantum systems would yield much higher fidelities. We are
presently implementing a multiple-round HBAC algorithm with the goal
of moving beyond the 1.5 signal amplification benchmark that uses only
unitary transformation.

\section{Control of coupled electron-nuclear systems}
\label{4}

\begin{figure}
\scalebox{0.25}{\includegraphics{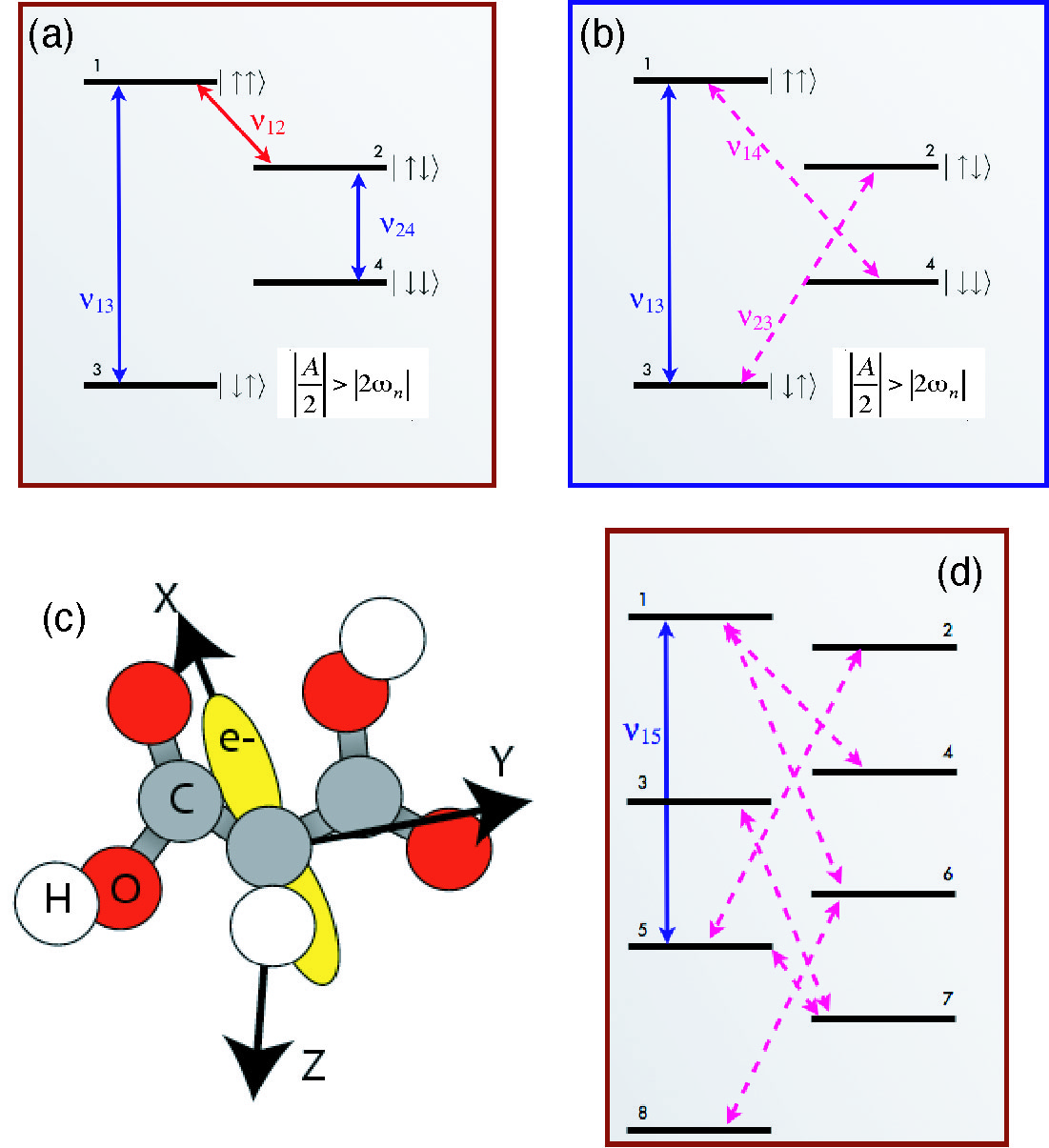}}
\caption{(a) The ENDOR model for universal quantum control of an
  electron-nuclear hyperfine coupled spin pair.  All states can be
  reached using three control parameters: two electron transitions at
  microwave frequencies (blue arrows) and one nuclear transition at
  radio-frequency (red arrow). (b) If the system has an anisotropic
  hyperfine interaction, universal control can be achieved using only
  one microwave control field.  The mixing of states connected by
  ``forbidden'' transitions (dashed arrows) occurs under the hyperfine
  interaction, allowing pathways to all states.  Modulated sequences
  of the single control field can be found using the GRAPE method to
  generate arbitrary unitary gates.  (c) Stable radical of malonic
  acid, produced by x-ray irradiation.  One of the methylene protons
  is removed by irradiation, leaving behind an unpaired $\pi$-electron
  spin-1/2.  The maximum hyperfine couplings to the methylene $^{13}C$
  and $^{1}H$ are $220$ and $90$ MHz, respectively.  (d) A
  representative level connectivity diagram for a 1e-2n system in an
  orientation for which the anisotropic hyperfine terms are large.
  Universal quantum control is achievable by implementing appropriate
  modulated sequences on the microwave transition labeled $\nu_{15}$.
  A $^{13}C$ labeled nucleus on the methylene carbon of malonic acid
  would exhibit a similar level structure.}
\label{fig:hyper}
\end{figure}

An attractive direction for spin-based quantum information processing
is to couple the nuclear spin degree of freedom with that of the
electron spin.  The larger magnetic moment of the electron leads to a
much faster timescale for spin manipulation, and the electron's charge
degree of freedom can serve as a pathway for initialization and
readout of spin qubits.  Early progress in this direction was made by
Mehring {\it et al.} \cite{MMS03a,MSW04a}, who showed coherent control of an
ensemble of electron-proton spin pairs in irradiated malonic acid
using electron-nuclear double resonance (ENDOR) techniques.  This
control scheme is sketched in Figure~\ref{fig:hyper}a, where universal
quantum control is achieved in principle by pulse sequences involving
two microwave and one RF field.  

Moreover, Mehring introduced the important idea that, in hyperfine
systems with one electron and $N>1$ distinguishable nuclear spins,
quantum gates operating on nuclear states could be sped up
significantly using the hyperfine interaction with the electron to
mediate indirect nuclear-nuclear interactions \cite{MM06a}.  This was
dubbed the S-bus concept since the electron spin operator is often
denoted as $\bf{S}$.  The S-bus concept is suggestive of a picture in
which isolated clusters of 1 electron + $N$ nuclear spins form qubit
sub-registers, where a handful of qubits can be manipulated and stored
with high fidelity.

Initialization of the electron spins to the ground state $\ket{0}$
occurs thermally at modestly low temperatures and typical external
magnetic fields.  The nuclear spin qubits can be initialized one by
one by alternating polarization swaps with the electron spin and
waiting for subsequent electron thermalization
($T^{elec}_1<<T^{nuc}_1$). There are various possibilities for
coupling sub-registers: the electron-electron exchange coupling
modulated by electrostatic gates \cite{Kan98a}, direct dipolar
coupling between electrons, or long-range coupling using optical
techniques \cite{JTS+07a,JGP+04a,JGP+04b,CDT+06a}. A recent proposal
outlines a method to further improve the computational power of the
S-bus concept.  This proposed method uses the natural anisotropy of
the hyperfine interaction to ``mix'' spin sublevels (much as the
strong coupling effect discussed above mixes nuclear spin
computational basis states). This eliminates the need for direct RF
control of the nuclear spin transitions. This method is shown
schematically in Figure~\ref{fig:hyper}b. The effective spin
Hamiltonian of the electron-nuclear pair has the form:
\begin{equation}
\ham_{en}=\omega_e^LS_z + \omega_n^LI_{z} + A_zS_zI_z+A_xS_zI_x, 
\end{equation}
where $A_x$ ($A_z$) is the anisotropic (isoptropic) hyperfine
coupling, $\bf{S}$ and $\bf{I}$ are the electron and nuclear spin
operators, and $\omega_{e,n}^L$ are the Larmor frequencies.  If we
arrange the system such that $A_x\sim\omega_n^L$, there will be
efficient mixing of the sub-levels indicated by the dashed arrows in
figure~\ref{fig:hyper}b, due to the hyperfine term $A_xS_zI_x$.

It can be shown mathematically that controlled driving of a single
microwave transition (indicated as the 1-3 transition in
figure~\ref{fig:hyper}b) is enough to generate any unitary operator
when given sufficient bandwidth, duration and complexity of the
control field.  This remarkable result
allows for much faster operations by replacing the low energy nuclear
excitation field with the higher energy hyperfine coupling.  It also
simplifies the experimental setup by requiring only one control
frequency. Moreover, we can apply optimal control pulse-shaping
methods such as GRAPE to generate universal quantum gates.
Experimental work is currently underway with the aim of demonstrating
these techniques using hyperfine coupled spin systems such as the
stable malonic acid radical (see figure~\ref{fig:hyper}c-d).

\section{Conclusion}
\label{5}

Of the many proposed and current realizations of quantum information
processing devices, NMR stands out as the present leader in terms of
control fidelities and the manipulation of the largest number of
quantum bits. This has allowed for the implementation of a variety of
benchmarks and algorithms that bring theoretical concepts to the
laboratory.  Its success is built on many years of RF engineering
invested in NMR technology for a variety of purposes that have
recently been adapted for quantum information processing. Combining
such control with the long coherence times of nuclear spins makes NMR
promising for quantum computation.

One of the present difficulties with using only nuclear spins is the
ability to initialize them into pure states.  Therefore the next
generation of magnetic resonance based devices will make use of
electron spins both to initialize nuclear spin qubits and to mediate
interactions between them, yielding much faster gate operations. We
expect this research to continue to produce exciting results and to
merge with other quantum technologies in the coming years.

\begin{acknowledgments}
  We would like to thank NSERC and CIAR support.  This work was
  also supported in part by the National Security Agency (NSA) under Army
  Research Office (ARO) contract numbers W911NF-05-1-0469 and
  DAAD19-01-1-0519, by the Air Force Office of Scientific Research,
  and by the Quantum Technologies Group of the Cambridge-MIT
  Institute.
\end{acknowledgments}

\bibliography{qubibcap}

\begin{thebibliography}{38}
\expandafter\ifx\csname natexlab\endcsname\relax\def\natexlab#1{#1}\fi
\expandafter\ifx\csname bibnamefont\endcsname\relax
  \def\bibnamefont#1{#1}\fi
\expandafter\ifx\csname bibfnamefont\endcsname\relax
  \def\bibfnamefont#1{#1}\fi
\expandafter\ifx\csname citenamefont\endcsname\relax
  \def\citenamefont#1{#1}\fi
\expandafter\ifx\csname url\endcsname\relax
  \def\url#1{\texttt{#1}}\fi
\expandafter\ifx\csname urlprefix\endcsname\relax\def\urlprefix{URL }\fi
\providecommand{\bibinfo}[2]{#2}
\providecommand{\eprint}[2][]{\url{#2}}

\bibitem[{\citenamefont{Purcell et~al.}(1946)\citenamefont{Purcell, Torrey, and
  Pound}}]{PTP45a}
\bibinfo{author}{\bibfnamefont{E.~M.} \bibnamefont{Purcell}},
  \bibinfo{author}{\bibfnamefont{H.~C.} \bibnamefont{Torrey}},
  \bibnamefont{and} \bibinfo{author}{\bibfnamefont{R.~V.} \bibnamefont{Pound}},
  \bibinfo{journal}{Phys. Rev.} \textbf{\bibinfo{volume}{69}},
  \bibinfo{pages}{37} (\bibinfo{year}{1946}).

\bibitem[{\citenamefont{Bloch}(1946)}]{Blo46a}
\bibinfo{author}{\bibfnamefont{F.}~\bibnamefont{Bloch}},
  \bibinfo{journal}{Phys. Rev.} \textbf{\bibinfo{volume}{70}},
  \bibinfo{pages}{460} (\bibinfo{year}{1946}).

\bibitem[{\citenamefont{Kaye et~al.}(2007)\citenamefont{Kaye, Laflamme, and
  Mosca}}]{KLM07a}
\bibinfo{author}{\bibfnamefont{P.}~\bibnamefont{Kaye}},
  \bibinfo{author}{\bibfnamefont{R.}~\bibnamefont{Laflamme}}, \bibnamefont{and}
  \bibinfo{author}{\bibfnamefont{M.}~\bibnamefont{Mosca}},
  \emph{\bibinfo{title}{An introduction to quantum computing}}
  (\bibinfo{publisher}{Oxford University Press}, \bibinfo{year}{2007}).

\bibitem[{\citenamefont{Levitt}(2001)}]{Lev01b}
\bibinfo{author}{\bibfnamefont{M.~H.} \bibnamefont{Levitt}},
  \emph{\bibinfo{title}{Spin dynamics: Basics of nuclear magnetic resonance}}
  (\bibinfo{publisher}{John Wiley {\&} Sons}, \bibinfo{address}{New-York},
  \bibinfo{year}{2001}).

\bibitem[{\citenamefont{Bowdrey et~al.}(2005)\citenamefont{Bowdrey, Jones,
  Knill, and Laflamme}}]{BJK+05a}
\bibinfo{author}{\bibfnamefont{M.~D.} \bibnamefont{Bowdrey}},
  \bibinfo{author}{\bibfnamefont{J.~A.} \bibnamefont{Jones}},
  \bibinfo{author}{\bibfnamefont{E.}~\bibnamefont{Knill}}, \bibnamefont{and}
  \bibinfo{author}{\bibfnamefont{R.}~\bibnamefont{Laflamme}},
  \bibinfo{journal}{Phys. Rev. A} \textbf{\bibinfo{volume}{72}},
  \bibinfo{eid}{032315} (\bibinfo{year}{2005}).

\bibitem[{\citenamefont{Fortunato et~al.}(2002)\citenamefont{Fortunato, Pravia,
  Boulant, Teklemariam, Havel, and Cory}}]{FPB+02a}
\bibinfo{author}{\bibfnamefont{E.~M.} \bibnamefont{Fortunato}},
  \bibinfo{author}{\bibfnamefont{M.~A.} \bibnamefont{Pravia}},
  \bibinfo{author}{\bibfnamefont{N.}~\bibnamefont{Boulant}},
  \bibinfo{author}{\bibfnamefont{G.}~\bibnamefont{Teklemariam}},
  \bibinfo{author}{\bibfnamefont{T.~F.} \bibnamefont{Havel}}, \bibnamefont{and}
  \bibinfo{author}{\bibfnamefont{D.~G.} \bibnamefont{Cory}},
  \bibinfo{journal}{J. Chem. Phys.} \textbf{\bibinfo{volume}{116}},
  \bibinfo{pages}{7599} (\bibinfo{year}{2002}).

\bibitem[{\citenamefont{Khaneja et~al.}(2005)\citenamefont{Khaneja, Reiss,
  Kehlet, Schulte-Herbruggen, and Glaser}}]{KRK+05a}
\bibinfo{author}{\bibfnamefont{N.}~\bibnamefont{Khaneja}},
  \bibinfo{author}{\bibfnamefont{T.}~\bibnamefont{Reiss}},
  \bibinfo{author}{\bibfnamefont{C.}~\bibnamefont{Kehlet}},
  \bibinfo{author}{\bibfnamefont{T.}~\bibnamefont{Schulte-Herbruggen}},
  \bibnamefont{and} \bibinfo{author}{\bibfnamefont{S.~J.}
  \bibnamefont{Glaser}}, \bibinfo{journal}{J. Mag. Res.}
  \textbf{\bibinfo{volume}{172}}, \bibinfo{pages}{296} (\bibinfo{year}{2005}).

\bibitem[{\citenamefont{Rabitz et~al.}(2000)\citenamefont{Rabitz,
  de~Vivie-Riedle, Motzkus, and Kompa}}]{RVM+00a}
\bibinfo{author}{\bibfnamefont{H.}~\bibnamefont{Rabitz}},
  \bibinfo{author}{\bibfnamefont{R.}~\bibnamefont{de~Vivie-Riedle}},
  \bibinfo{author}{\bibfnamefont{M.}~\bibnamefont{Motzkus}}, \bibnamefont{and}
  \bibinfo{author}{\bibfnamefont{K.}~\bibnamefont{Kompa}},
  \bibinfo{journal}{Science} \textbf{\bibinfo{volume}{288}},
  \bibinfo{pages}{824} (\bibinfo{year}{2000}).

\bibitem[{\citenamefont{Bryson and Ho}(1975)}]{BH75a}
\bibinfo{author}{\bibfnamefont{A.~E.} \bibnamefont{Bryson}} \bibnamefont{and}
  \bibinfo{author}{\bibfnamefont{Y.~C.} \bibnamefont{Ho}},
  \emph{\bibinfo{title}{Applied optimal control: optimization, estimation, and
  control}} (\bibinfo{publisher}{Hemisphere Pub. Corp. Washington DC},
  \bibinfo{year}{1975}).

\bibitem[{\citenamefont{Negrevergne et~al.}(2006)\citenamefont{Negrevergne,
  Mahesh, Ryan, Ditty, Cyr-Racine, Power, Boulant, Havel, Cory, and
  Laflamme}}]{NMR+06a}
\bibinfo{author}{\bibfnamefont{C.}~\bibnamefont{Negrevergne}},
  \bibinfo{author}{\bibfnamefont{T.~S.} \bibnamefont{Mahesh}},
  \bibinfo{author}{\bibfnamefont{C.~A.} \bibnamefont{Ryan}},
  \bibinfo{author}{\bibfnamefont{M.~J.} \bibnamefont{Ditty}},
  \bibinfo{author}{\bibfnamefont{F.}~\bibnamefont{Cyr-Racine}},
  \bibinfo{author}{\bibfnamefont{W.}~\bibnamefont{Power}},
  \bibinfo{author}{\bibfnamefont{N.}~\bibnamefont{Boulant}},
  \bibinfo{author}{\bibfnamefont{T.~F.} \bibnamefont{Havel}},
  \bibinfo{author}{\bibfnamefont{D.~G.} \bibnamefont{Cory}}, \bibnamefont{and}
  \bibinfo{author}{\bibfnamefont{R.}~\bibnamefont{Laflamme}},
  \bibinfo{journal}{Phys. Rev. Lett.} \textbf{\bibinfo{volume}{96}},
  \bibinfo{pages}{17501} (\bibinfo{year}{2006}).

\bibitem[{\citenamefont{Cory et~al.}(1998)\citenamefont{Cory, Price, Maas,
  Knill, Laflamme, Zurek, Havel, and Somaroo}}]{CPM+98a}
\bibinfo{author}{\bibfnamefont{D.~G.} \bibnamefont{Cory}},
  \bibinfo{author}{\bibfnamefont{M.~D.} \bibnamefont{Price}},
  \bibinfo{author}{\bibfnamefont{W.}~\bibnamefont{Maas}},
  \bibinfo{author}{\bibfnamefont{E.}~\bibnamefont{Knill}},
  \bibinfo{author}{\bibfnamefont{R.}~\bibnamefont{Laflamme}},
  \bibinfo{author}{\bibfnamefont{W.~H.} \bibnamefont{Zurek}},
  \bibinfo{author}{\bibfnamefont{T.~F.} \bibnamefont{Havel}}, \bibnamefont{and}
  \bibinfo{author}{\bibfnamefont{S.~S.} \bibnamefont{Somaroo}},
  \bibinfo{journal}{Phys. Rev. Lett.} \textbf{\bibinfo{volume}{81}},
  \bibinfo{pages}{2152} (\bibinfo{year}{1998}).

\bibitem[{\citenamefont{Laforest et~al.}(2007)\citenamefont{Laforest, Simon,
  Boileau, Baugh, Ditty, and Laflamme}}]{LSB+07a}
\bibinfo{author}{\bibfnamefont{M.}~\bibnamefont{Laforest}},
  \bibinfo{author}{\bibfnamefont{D.}~\bibnamefont{Simon}},
  \bibinfo{author}{\bibfnamefont{J.~C.} \bibnamefont{Boileau}},
  \bibinfo{author}{\bibfnamefont{J.}~\bibnamefont{Baugh}},
  \bibinfo{author}{\bibfnamefont{M.~J.} \bibnamefont{Ditty}}, \bibnamefont{and}
  \bibinfo{author}{\bibfnamefont{R.}~\bibnamefont{Laflamme}},
  \bibinfo{journal}{Phys. Rev. A} \textbf{\bibinfo{volume}{75}},
  \bibinfo{eid}{012331} (\bibinfo{year}{2007}).

\bibitem[{\citenamefont{Cory et~al.}(2000)\citenamefont{Cory, Somaroo, Havel,
  Knill, Laflamme, and Zurek}}]{CSH+00a}
\bibinfo{author}{\bibfnamefont{D.~G.} \bibnamefont{Cory}},
  \bibinfo{author}{\bibfnamefont{S.~S.} \bibnamefont{Somaroo}},
  \bibinfo{author}{\bibfnamefont{T.~F.} \bibnamefont{Havel}},
  \bibinfo{author}{\bibfnamefont{E.}~\bibnamefont{Knill}},
  \bibinfo{author}{\bibfnamefont{R.}~\bibnamefont{Laflamme}}, \bibnamefont{and}
  \bibinfo{author}{\bibfnamefont{W.~H.} \bibnamefont{Zurek}},
  \bibinfo{journal}{Mol. Phys.} \textbf{\bibinfo{volume}{98}},
  \bibinfo{pages}{1347} (\bibinfo{year}{2000}).

\bibitem[{\citenamefont{Knill et~al.}(2001)\citenamefont{Knill, Laflamme,
  Martinez, and Negrevergne}}]{KLM+01a}
\bibinfo{author}{\bibfnamefont{E.}~\bibnamefont{Knill}},
  \bibinfo{author}{\bibfnamefont{R.}~\bibnamefont{Laflamme}},
  \bibinfo{author}{\bibfnamefont{R.}~\bibnamefont{Martinez}}, \bibnamefont{and}
  \bibinfo{author}{\bibfnamefont{C.}~\bibnamefont{Negrevergne}},
  \emph{\bibinfo{title}{Implementation of the five qubit error correction
  benchmark}} (\bibinfo{year}{2001}), \eprint{quant-ph/0101034}.

\bibitem[{\citenamefont{Boulant et~al.}(2004)\citenamefont{Boulant, Viola,
  Fortunato, and Cory}}]{BVF+04a}
\bibinfo{author}{\bibfnamefont{N.}~\bibnamefont{Boulant}},
  \bibinfo{author}{\bibfnamefont{L.}~\bibnamefont{Viola}},
  \bibinfo{author}{\bibfnamefont{E.~M.} \bibnamefont{Fortunato}},
  \bibnamefont{and} \bibinfo{author}{\bibfnamefont{D.~G.} \bibnamefont{Cory}},
  \emph{\bibinfo{title}{Experimental implementation of a concatenated quantum
  error-correcting code}} (\bibinfo{year}{2004}), \eprint{quant-ph/0409193}.

\bibitem[{\citenamefont{Chen et~al.}(2004)\citenamefont{Chen, Yepez, and
  Cory}}]{CYC04a}
\bibinfo{author}{\bibfnamefont{Z.}~\bibnamefont{Chen}},
  \bibinfo{author}{\bibfnamefont{J.}~\bibnamefont{Yepez}}, \bibnamefont{and}
  \bibinfo{author}{\bibfnamefont{D.~G.} \bibnamefont{Cory}},
  \emph{\bibinfo{title}{Simulation of the burgers equation by nmr quantum
  information processing}} (\bibinfo{year}{2004}), \eprint{quant-ph/0410198}.

\bibitem[{\citenamefont{Negrevergne et~al.}(2004)\citenamefont{Negrevergne,
  Somma, Ortiz, Knill, and Laflamme}}]{NSO+04a}
\bibinfo{author}{\bibfnamefont{C.}~\bibnamefont{Negrevergne}},
  \bibinfo{author}{\bibfnamefont{R.}~\bibnamefont{Somma}},
  \bibinfo{author}{\bibfnamefont{G.}~\bibnamefont{Ortiz}},
  \bibinfo{author}{\bibfnamefont{E.}~\bibnamefont{Knill}}, \bibnamefont{and}
  \bibinfo{author}{\bibfnamefont{R.}~\bibnamefont{Laflamme}},
  \emph{\bibinfo{title}{Liquid state nmr simulations of quantum many-body
  problems}} (\bibinfo{year}{2004}), \eprint{quant-ph/0410106}.

\bibitem[{\citenamefont{Havel et~al.}(2001)\citenamefont{Havel, Sharf, Viola,
  and Cory}}]{HSV+01a}
\bibinfo{author}{\bibfnamefont{T.~F.} \bibnamefont{Havel}},
  \bibinfo{author}{\bibfnamefont{Y.}~\bibnamefont{Sharf}},
  \bibinfo{author}{\bibfnamefont{L.}~\bibnamefont{Viola}}, \bibnamefont{and}
  \bibinfo{author}{\bibfnamefont{D.~G.} \bibnamefont{Cory}},
  \bibinfo{journal}{Phys. Rev. A} \textbf{\bibinfo{volume}{vol. 280}},
  \bibinfo{pages}{282} (\bibinfo{year}{2001}).

\bibitem[{\citenamefont{Tseng et~al.}(1999)\citenamefont{Tseng, Somaroo, Sharf,
  Knill, Laflamme, Havel, and Cory}}]{TSS+99a}
\bibinfo{author}{\bibfnamefont{C.~H.} \bibnamefont{Tseng}},
  \bibinfo{author}{\bibfnamefont{S.}~\bibnamefont{Somaroo}},
  \bibinfo{author}{\bibfnamefont{Y.}~\bibnamefont{Sharf}},
  \bibinfo{author}{\bibfnamefont{E.}~\bibnamefont{Knill}},
  \bibinfo{author}{\bibfnamefont{R.}~\bibnamefont{Laflamme}},
  \bibinfo{author}{\bibfnamefont{T.~F.} \bibnamefont{Havel}}, \bibnamefont{and}
  \bibinfo{author}{\bibfnamefont{D.~G.} \bibnamefont{Cory}},
  \emph{\bibinfo{title}{Quantum simulation of a three-body interaction
  hamiltonian on an nmr quantum computer}} (\bibinfo{year}{1999}),
  \eprint{quant-ph/9908012}.

\bibitem[{\citenamefont{Somaroo et~al.}(1999)\citenamefont{Somaroo, Tseng,
  Havel, Laflamme, and Cory}}]{STH+99a}
\bibinfo{author}{\bibfnamefont{S.}~\bibnamefont{Somaroo}},
  \bibinfo{author}{\bibfnamefont{C.~H.} \bibnamefont{Tseng}},
  \bibinfo{author}{\bibfnamefont{T.~F.} \bibnamefont{Havel}},
  \bibinfo{author}{\bibfnamefont{R.}~\bibnamefont{Laflamme}}, \bibnamefont{and}
  \bibinfo{author}{\bibfnamefont{D.~G.} \bibnamefont{Cory}},
  \bibinfo{journal}{Phys. Rev. Lett.} \textbf{\bibinfo{volume}{82}},
  \bibinfo{pages}{5381} (\bibinfo{year}{1999}).

\bibitem[{\citenamefont{Abragam and Goldman}(1982)}]{AG82a}
\bibinfo{author}{\bibfnamefont{A.}~\bibnamefont{Abragam}} \bibnamefont{and}
  \bibinfo{author}{\bibfnamefont{M.}~\bibnamefont{Goldman}},
  \emph{\bibinfo{title}{Nuclear Magnetism: Order and Disorder}}
  (\bibinfo{publisher}{Oxford Scientific Publishers}, \bibinfo{address}{New
  York}, \bibinfo{year}{1982}).

\bibitem[{\citenamefont{Kane}(1998)}]{Kan98a}
\bibinfo{author}{\bibfnamefont{B.~E.} \bibnamefont{Kane}},
  \bibinfo{journal}{Nature} \textbf{\bibinfo{volume}{393}},
  \bibinfo{pages}{133} (\bibinfo{year}{1998}).

\bibitem[{\citenamefont{Jiang et~al.}(2007)\citenamefont{Jiang, Taylor,
  Sorenson, and Lukin}}]{JTS+07a}
\bibinfo{author}{\bibfnamefont{L.}~\bibnamefont{Jiang}},
  \bibinfo{author}{\bibfnamefont{J.~M.} \bibnamefont{Taylor}},
  \bibinfo{author}{\bibfnamefont{A.~S.} \bibnamefont{Sorenson}},
  \bibnamefont{and} \bibinfo{author}{\bibfnamefont{M.~D.} \bibnamefont{Lukin}}
  (\bibinfo{year}{2007}), \eprint{quant-ph/0703029}.

\bibitem[{\citenamefont{Mehring and Mende}(2006)}]{MM06a}
\bibinfo{author}{\bibfnamefont{M.}~\bibnamefont{Mehring}} \bibnamefont{and}
  \bibinfo{author}{\bibfnamefont{J.}~\bibnamefont{Mende}},
  \bibinfo{journal}{Phys. Rev. A} \textbf{\bibinfo{volume}{73}},
  \bibinfo{eid}{052303} (\bibinfo{year}{2006}).

\bibitem[{\citenamefont{Jelezko
  et~al.}(2004{\natexlab{a}})\citenamefont{Jelezko, Gaebel, Popa, Gruber, and
  Wrachtrup}}]{JGP+04a}
\bibinfo{author}{\bibfnamefont{F.}~\bibnamefont{Jelezko}},
  \bibinfo{author}{\bibfnamefont{T.}~\bibnamefont{Gaebel}},
  \bibinfo{author}{\bibfnamefont{I.}~\bibnamefont{Popa}},
  \bibinfo{author}{\bibfnamefont{A.}~\bibnamefont{Gruber}}, \bibnamefont{and}
  \bibinfo{author}{\bibfnamefont{J.}~\bibnamefont{Wrachtrup}},
  \bibinfo{journal}{Phys. Rev. Lett.} \textbf{\bibinfo{volume}{92}},
  \bibinfo{eid}{076401} (\bibinfo{year}{2004}{\natexlab{a}}).

\bibitem[{\citenamefont{Jelezko
  et~al.}(2004{\natexlab{b}})\citenamefont{Jelezko, Gaebel, Popa, Domhan,
  Gruber, and Wrachtrup}}]{JGP+04b}
\bibinfo{author}{\bibfnamefont{F.}~\bibnamefont{Jelezko}},
  \bibinfo{author}{\bibfnamefont{T.}~\bibnamefont{Gaebel}},
  \bibinfo{author}{\bibfnamefont{I.}~\bibnamefont{Popa}},
  \bibinfo{author}{\bibfnamefont{M.}~\bibnamefont{Domhan}},
  \bibinfo{author}{\bibfnamefont{A.}~\bibnamefont{Gruber}}, \bibnamefont{and}
  \bibinfo{author}{\bibfnamefont{J.}~\bibnamefont{Wrachtrup}},
  \bibinfo{journal}{Phys. Rev. Lett.} \textbf{\bibinfo{volume}{93}},
  \bibinfo{eid}{130501} (\bibinfo{year}{2004}{\natexlab{b}}).

\bibitem[{\citenamefont{Childress et~al.}(2006)\citenamefont{Childress,
  Gurudev~Dutt, Taylor, Zibrov, Jelezko, Wrachtrup, Hemmer, and
  Lukin}}]{CDT+06a}
\bibinfo{author}{\bibfnamefont{L.}~\bibnamefont{Childress}},
  \bibinfo{author}{\bibfnamefont{M.~V.} \bibnamefont{Gurudev~Dutt}},
  \bibinfo{author}{\bibfnamefont{J.~M.} \bibnamefont{Taylor}},
  \bibinfo{author}{\bibfnamefont{A.~S.} \bibnamefont{Zibrov}},
  \bibinfo{author}{\bibfnamefont{F.}~\bibnamefont{Jelezko}},
  \bibinfo{author}{\bibfnamefont{J.}~\bibnamefont{Wrachtrup}},
  \bibinfo{author}{\bibfnamefont{P.~R.} \bibnamefont{Hemmer}},
  \bibnamefont{and} \bibinfo{author}{\bibfnamefont{M.~D.} \bibnamefont{Lukin}},
  \bibinfo{journal}{Science} \textbf{\bibinfo{volume}{314}},
  \bibinfo{pages}{281} (\bibinfo{year}{2006}).

\bibitem[{\citenamefont{McCalley and Kwiram}(1993)}]{MK93a}
\bibinfo{author}{\bibfnamefont{R.~C.} \bibnamefont{McCalley}} \bibnamefont{and}
  \bibinfo{author}{\bibfnamefont{A.~L.} \bibnamefont{Kwiram}},
  \bibinfo{journal}{J. Phys. Chem.} \textbf{\bibinfo{volume}{97}},
  \bibinfo{pages}{2888} (\bibinfo{year}{1993}).

\bibitem[{\citenamefont{Baugh et~al.}(2006)\citenamefont{Baugh, Moussa, Ryan,
  Laflamme, Ramanathan, Havel, and Cory}}]{BMR+06a}
\bibinfo{author}{\bibfnamefont{J.}~\bibnamefont{Baugh}},
  \bibinfo{author}{\bibfnamefont{O.}~\bibnamefont{Moussa}},
  \bibinfo{author}{\bibfnamefont{C.~A.} \bibnamefont{Ryan}},
  \bibinfo{author}{\bibfnamefont{R.}~\bibnamefont{Laflamme}},
  \bibinfo{author}{\bibfnamefont{C.}~\bibnamefont{Ramanathan}},
  \bibinfo{author}{\bibfnamefont{T.~F.} \bibnamefont{Havel}}, \bibnamefont{and}
  \bibinfo{author}{\bibfnamefont{D.~G.} \bibnamefont{Cory}},
  \bibinfo{journal}{Phys. Rev. A} \textbf{\bibinfo{volume}{73}},
  \bibinfo{eid}{022305} (\bibinfo{year}{2006}).

\bibitem[{\citenamefont{Toffoli}(1980)}]{Tof80a}
\bibinfo{author}{\bibfnamefont{T.}~\bibnamefont{Toffoli}}, in
  \emph{\bibinfo{booktitle}{Automata, Languages and Programming}}, edited by
  \bibinfo{editor}{\bibfnamefont{W.}~\bibnamefont{de~Bakker}} \bibnamefont{and}
  \bibinfo{editor}{\bibfnamefont{J.}~\bibnamefont{van Leeuwen}}
  (\bibinfo{publisher}{Springer}, \bibinfo{address}{New York},
  \bibinfo{year}{1980}), p. \bibinfo{pages}{632}, \bibinfo{note}{technical Memo
  MIT/LCS/TM-151, MIT Lab. for Comput. Sci. (unpublished).}

\bibitem[{\citenamefont{Baugh et~al.}(2005)\citenamefont{Baugh, Moussa, Ryan,
  Nayak, and Laflamme}}]{BMR+05a}
\bibinfo{author}{\bibfnamefont{J.}~\bibnamefont{Baugh}},
  \bibinfo{author}{\bibfnamefont{O.}~\bibnamefont{Moussa}},
  \bibinfo{author}{\bibfnamefont{C.~A.} \bibnamefont{Ryan}},
  \bibinfo{author}{\bibfnamefont{A.}~\bibnamefont{Nayak}}, \bibnamefont{and}
  \bibinfo{author}{\bibfnamefont{R.}~\bibnamefont{Laflamme}},
  \bibinfo{journal}{Nature} \textbf{\bibinfo{volume}{438}},
  \bibinfo{pages}{470} (\bibinfo{year}{2005}).

\bibitem[{\citenamefont{Schulman et~al.}(2005)\citenamefont{Schulman, Mor, and
  Weinstein}}]{SMW05a}
\bibinfo{author}{\bibfnamefont{L.}~\bibnamefont{Schulman}},
  \bibinfo{author}{\bibfnamefont{T.}~\bibnamefont{Mor}}, \bibnamefont{and}
  \bibinfo{author}{\bibfnamefont{Y.}~\bibnamefont{Weinstein}},
  \bibinfo{journal}{Phys. Rev. Lett.} \textbf{\bibinfo{volume}{94}},
  \bibinfo{pages}{120501} (\bibinfo{year}{2005}).

\bibitem[{\citenamefont{Boykin et~al.}(2002)\citenamefont{Boykin, Mor,
  Roychowdhury, Vatan, and Vrijen}}]{BMR+02a}
\bibinfo{author}{\bibfnamefont{P.~O.} \bibnamefont{Boykin}},
  \bibinfo{author}{\bibfnamefont{T.}~\bibnamefont{Mor}},
  \bibinfo{author}{\bibfnamefont{V.}~\bibnamefont{Roychowdhury}},
  \bibinfo{author}{\bibfnamefont{F.}~\bibnamefont{Vatan}}, \bibnamefont{and}
  \bibinfo{author}{\bibfnamefont{R.}~\bibnamefont{Vrijen}},
  \bibinfo{journal}{Proc. Natl. Acad. Sci. USA} \textbf{\bibinfo{volume}{99}},
  \bibinfo{pages}{3388} (\bibinfo{year}{2002}).

\bibitem[{\citenamefont{Weitekamp et~al.}(1982)\citenamefont{Weitekamp, Garbow,
  and Pines}}]{WGP82a}
\bibinfo{author}{\bibfnamefont{D.~P.} \bibnamefont{Weitekamp}},
  \bibinfo{author}{\bibfnamefont{J.~R.} \bibnamefont{Garbow}},
  \bibnamefont{and} \bibinfo{author}{\bibfnamefont{A.}~\bibnamefont{Pines}},
  \bibinfo{journal}{J. Chem. Phys.} \textbf{\bibinfo{volume}{77}},
  \bibinfo{pages}{2870} (\bibinfo{year}{1982}).

\bibitem[{\citenamefont{Hartmann and Hahn}(1962)}]{HH62a}
\bibinfo{author}{\bibfnamefont{S.~R.} \bibnamefont{Hartmann}} \bibnamefont{and}
  \bibinfo{author}{\bibfnamefont{E.~L.} \bibnamefont{Hahn}},
  \bibinfo{journal}{Phys. Rev.} \textbf{\bibinfo{volume}{128}},
  \bibinfo{pages}{2042} (\bibinfo{year}{1962}).

\bibitem[{\citenamefont{Zhang and Cory}(1998)}]{ZC98a}
\bibinfo{author}{\bibfnamefont{W.}~\bibnamefont{Zhang}} \bibnamefont{and}
  \bibinfo{author}{\bibfnamefont{D.~G.} \bibnamefont{Cory}},
  \bibinfo{journal}{Phys. Rev. Lett.} \textbf{\bibinfo{volume}{80}},
  \bibinfo{pages}{1324} (\bibinfo{year}{1998}).

\bibitem[{\citenamefont{Mehring et~al.}(2003)\citenamefont{Mehring, Mende, and
  Scherer}}]{MMS03a}
\bibinfo{author}{\bibfnamefont{M.}~\bibnamefont{Mehring}},
  \bibinfo{author}{\bibfnamefont{J.}~\bibnamefont{Mende}}, \bibnamefont{and}
  \bibinfo{author}{\bibfnamefont{W.}~\bibnamefont{Scherer}},
  \bibinfo{journal}{Phys. Rev. Lett.} \textbf{\bibinfo{volume}{90}},
  \bibinfo{pages}{153001} (\bibinfo{year}{2003}).

\bibitem[{\citenamefont{Mehring et~al.}(2004)\citenamefont{Mehring, Scherer,
  and Weidinger}}]{MSW04a}
\bibinfo{author}{\bibfnamefont{M.}~\bibnamefont{Mehring}},
  \bibinfo{author}{\bibfnamefont{W.}~\bibnamefont{Scherer}}, \bibnamefont{and}
  \bibinfo{author}{\bibfnamefont{A.}~\bibnamefont{Weidinger}},
  \bibinfo{journal}{Phys. Rev. Lett.} \textbf{\bibinfo{volume}{93}},
  \bibinfo{eid}{206603} (\bibinfo{year}{2004}).

\end{thebibliography}

\end{document}